\documentclass[aps,pra,reprint]{revtex4-1}
\usepackage{graphicx}
\usepackage{dcolumn}
\usepackage{bm}
\usepackage{mdwlist}
\usepackage{setspace}
\usepackage{natbib}
\usepackage{amssymb}
\usepackage{amsmath}
\usepackage{paralist}
\usepackage{enumitem}
\usepackage{fouridx}
\usepackage{overpic}
\usepackage{cases}
\usepackage{rotating}
\usepackage{color}
\DeclareMathOperator{\re}{Re}
\DeclareMathOperator{\im}{Im}

\begin{document}

\title{Ponderomotive manipulation of cold subwavelength plasmas}

\author{P. W. Smorenburg}
\author{L. P. J. Kamp}
\author{O. J. Luiten}
\email{o.j.luiten@tue.nl}
\affiliation{Coherence and Quantum Technology (CQT), Eindhoven University of Technology, PO Box 513, 5600 MB Eindhoven, The Netherlands}

\date{\today}

\begin{abstract}
Ponderomotive forces (PFs) induced in cold subwavelength plasmas by an externally applied electromagnetic wave are studied analytically. To this end, the plasma is modeled as a sphere with a radially varying permittivity, and the internal electric fields are calculated by solving the macroscopic Maxwell equations using an expansion in Debye potentials. It is found that the PF is directed opposite to the plasma density gradient, similarly to large-scale plasmas. In case of a uniform density profile, a residual spherically symmetric compressive PF is found, suggesting possibilities for contactless ponderomotive manipulation of homogeneous subwavelength objects. The presence of a surface PF on discontinuous plasma boundaries is derived. This force is essential for a microscopic description of the radiation-plasma interaction consistent with momentum conservation. It is shown that the PF integrated over the plasma volume is equivalent to the radiation pressure exerted on the plasma by the incident wave. The concept of radiative acceleration of subwavelength plasmas, proposed earlier, is applied to ultracold plasmas. It is estimated that these plasmas may be accelerated to keV ion energies, resulting in a neutralized beam with a brightness comparable to that of current high-performance ion sources.\\
\end{abstract}

\maketitle

\section{Introduction \label{sec1}}
Finite-sized plasmas driven by electromagnetic radiation are the subject of active studies in various research fields. Ultracold plasmas \cite{Killian}, which are created by photo-ionization of a cloud of laser-cooled atoms, are an exotic example of such finite-sized plasmas. They consist of up to $10^{11}$ singly-ionized atoms, have an electron temperature of $T_e\sim10$ K, and electron densities of $n_e\sim10^{18}$ m$^{-3}$. Ultracold plasmas are routinely probed with RF and microwave fields, enabling the observation of phenomena such as plasma oscillations \cite{Kulin,Bergeson}, Tonks-Dattner resonances \cite{Fletcher}, and modes associated with nonneutral plasmas \cite{Twedt,Lyubonko}. These observations in turn yield valuable fundamental insights into the plasma dynamics in the ultracold regime.\\
Laser-irradiated nanoplasmas \cite{Krainov,Fennel} constitute another class of finite-sized plasmas driven by electromagnetic radiation. Laser-driven atomic clusters are utilized as novel sources of intense pulses of electrons \cite{Chen-electrons,Fukuda-electrons}, ions \cite{Fukuda-ions}, and extreme ultraviolet \cite{Mori,Ter-Avetisyan-EUV} and x-ray \cite{Chen-xray} radiation. Directional proton beams can be produced by laser-irradiation of dense sub-micrometer-sized plasmas created from water droplets \cite{Ter-Avetisyan-ions}. Because the plasma frequency $\omega_p\propto\sqrt{n_e}$ in nanoplasmas is a factor $\sim10^5$ higher than in ultracold plasmas, nanoplasmas are usually subjected to optical rather than RF and microwave radiation. It is interesting that, despite the vastly different parameter regimes, ultracold plasmas and nanoplasmas have in common that the plasma size is smaller than the typically applied wavelength $\lambda\sim2\pi c/\omega_p$.\\

As long as the fields driving a finite-sized plasma are not so large that the excursions of the oscillating plasma electrons become comparable to the plasma size, escape of electrons and the resulting subsequent Coulomb expansion of the plasma are relatively unimportant \cite{Kishimoto}. In this so-called ambipolar or quasi-neutral regime (usually at electric field strengths below 1 MV/m for microwave radiation or at laser intensities $I\lesssim 10^{20}$ W/m$^2$ for optical frequencies), the plasma dynamics can be described hydrodynamically. In the one-fluid plasma model \cite{Mulser}, the plasma dynamics is governed by two force density contributions. The first of these is the well-known hydrodynamic force density $-\nabla p$, with $p=n_ek_BT_e$ the plasma pressure and $k_B$ Boltzmann's constant, which is present regardless of whether or not an external field is applied. The other is the ponderomotive force density,
\begin{align}
\bm{f}=-n_e\nabla\frac{e^2E^2}{4m_e\omega^2}\equiv-n_e\nabla\phi_p,\label{1}
\end{align}
induced by the external field. Here, $e$ is the elementary charge, $E$ the electric field strength, $m_e$ the electron mass, $\omega$ the applied frequency, and $\phi_p$ is the so-called ponderomotive potential. The force Eq. (\ref{1}) was originally derived for single electrons in an inhomogeneous ac field \cite{Boot,Gaponov}, and later extended to plasma media on the basis of the plasma fluid equations \cite{Hora,Shearer}. The relative importance of the hydrodynamic and ponderomotive forces, as is expressed in the ratio $\eta\equiv\left|-\nabla p\right|/\left|\bm{f}\right|\sim k_BT_e/\phi_p$, differs in nanoplasmas and ultracold plasmas in the quasi-neutral regime. Nanoplasmas have a temperature of at least $\sim1$ keV, so that $\eta\gtrsim1$ in the quasi-neutral regime, and hydrodynamic forces play an important role. Nevertheless, is has been recognized that ponderomotive forces can significantly modify the plasma dynamics even at relatively low intensities of $I\sim10^{19}$ W/m$^2$ \cite{Milchberg}. This reflects the complicated macroscopic behavior of dense finite plasmas, in which the hydrodynamic and electromagnetic effects are intertwined and difficult to study separately. In contrast, hydrodynamic forces are very small in ultracold plasmas. For $\omega/2\pi=1$ GHz and $T_e=10$ K, the force ratio $\eta<10^{-2}$ already for field strengths $E>10$ kV/m. In moderately to strongly driven ultracold plasmas, therefore, hydrodynamic forces are negligible compared to ponderomotive forces, contrary to the case of laser-excited nanoplasmas. This makes ultracold plasmas unique systems that can exhibit ponderomotive effects that are obscured in high-density plasmas.\\

In this paper, we study the ponderomotive forces induced in a finite-sized plasma by an applied electromagnetic wave, which are particularly important in the dynamics of ultracold plasmas, but are relevant to finite-sized plasmas in general. We concentrate on the typical circumstance that the plasma is smaller than the applied wavelength. The plasma is modeled as a sphere with a radially varying permittivity, and the electric field distribution is calculated by solving the macroscopic Maxwell equations in terms of an expansion in Debye potentials. This approach is commonly used to study the far field scattering properties of finite objects \cite{Kerker,Tai1,Tai2,Wyatt,Bisbing,Sun,Hoang,Orlov}, with little attention for the electromagnetic fields inside the object. An exception is a recent calculation of resonance absorption in dense atomic clusters based on the internal fields \cite{Holkundkar}. Here, we apply the technique to describe the opto-mechanical forces induced by the applied wave in the plasma itself. In view of the compressibility of the plasma, these forces form an essential part of the interaction dynamics. The following properties are found. First, the ponderomotive force in the plasma bulk is directed outwards for natural profiles $dn_e/dr<0$ and inwards for 'inverted' profiles $dn_e/dr>0$, where $r$ is the radial coordinate. Although this is similar to well-studied large-scale plasmas \cite{Hora}, there are also differences due to the subwavelength character of the system. Moreover, we find a spherically symmetric compressive ponderomotive force even in case of a completely uniform density. The latter suggests possibilities for contactless ponderomotive manipulation of subwavelength objects, which is not limited to plasmas but extends to other media with a well-defined uniform density. Second, we show that sharp plasma boundaries give rise to a ponderomotive surface force in a addition to the volume force corresponding to Eq. (\ref{1}). This surface force proves to be an essential ingredient in a correct local description of the interaction of electromagnetic waves with subwavelength objects that is consistent with momentum conservation. Third, we consider the total ponderomotive force integrated over the plasma volume and show that it is equivalent to the radiation pressure exerted on the plasma by the incident wave. In the past, it has been proposed to accelerate subwavelength plasmas with this radiation pressure \cite{Veksler,Motz}. Here, we assess the feasibility of this scheme for ultracold plasmas. We estimate that these plasmas may be accelerated to keV ion energies thanks to their extremely low temperature and correspondingly weak tendency to expand.\\

This paper is organized as follows. In order to properly describe the effects mentioned above, the electromagnetic fields and ponderomotive forces in the plasma are first formulated analytically in general terms in sections \ref{sec2} and \ref{sec3} respectively. These sections therefore have a mathematical character. Explicit results for the derived ponderomotive forces are summarized in section \ref{sec3b}, Eqs. (\ref{32})-(\ref{34}). These results are subsequently applied to concrete examples of plasmas in sections \ref{sec4} to \ref{sec6}. In section \ref{sec4}, a plasma with uniform density is considered, the compressive ponderomotive force is found, and the role of the ponderomotive surface force in the radiation pressure on the plasma is elucidated. Section \ref{sec5} concerns numerically calculated ponderomotive forces in inhomogeneous plasmas, exhibiting distinct bulk and surface contributions. In section \ref{sec6}, radiative acceleration of ultracold plasmas is discussed. Section \ref{sec7} concludes the paper.
\section{Fields \label{sec2}}
\subsection{Expansion in Debye potentials \label{sec2a}}
We consider a collisionless, unmagnetized, spherical plasma with radius $b$, in interaction with an incident linearly polarized plane wave with electric field $\bm{E}_{ext}=E_0\bm{e}_x\exp(ikz-i\omega t)$ and magnetic field $c\bm{B}_{ext}=E_0\bm{e}_y\exp(ikz-i\omega t)$. In this section, we discuss the electromagnetic field distribution in such a plasma. As is well known, the plasma can be treated as a medium with relative permittivity $\epsilon=1-\omega_p^2/\omega^2$, where $\omega_p(r)=\sqrt{n_ee^2/(m_e\epsilon_0)}$ is the local plasma frequency associated with the electron density $n_e(r)$ at radius $r$, and $\epsilon_0$ is the vacuum permittivity. The electromagnetic fields in the plasma satisfy the source-free macroscopic Maxwell equations \cite{Panofsky}.
Solution of these equations is analogous to the classical Mie scattering problem \cite{Bohren}, generalized to an object with a radially varying permittivity $\epsilon=\epsilon(r)$. This generalization has been worked out previously \cite{Kerker,Tai1,Wyatt,Sun}; we reproduce the results here because we will use them frequently in the remainder of the paper. The fields in the region $r<b$ inside the plasma can be decomposed \cite{Panofsky} into an electric (transverse magnetic) part $(\bm{E},\bm{B})^e$ with $B^e_r=0$ and a magnetic (transverse electric) part $(\bm{E},\bm{B})^m$ with $E^m_r=0$. These fields can be written in terms of two scalar Debye potentials $\Pi^{e,m}(\bm{r})$ as
\begin{align}
\begin{array}{rlcrl}
\bm{E}^m&=E_0\bm{r}\times\nabla\Pi^m ;&\hphantom{=}&i\omega\bm{B}^m&=\nabla\times\bm{E}^m;\\
-c\bm{B}^e&=E_0\bm{r}\times\nabla\Pi^e;&\hphantom{=}&-i\omega\bm{D}^e&=\nabla\times\bm{H}^e,
\end{array}\label{2}
\end{align}
where $\bm{D}^e=\epsilon_0\epsilon\bm{E}^e$ and $\bm{H}^e=\bm{B}^e/\mu_0$ with $\mu_0$ the vacuum permeability, and factors $\exp(-i\omega t)$ have been suppressed. In spherical coordinates $(r,\theta,\varphi)$, the potentials evaluate to $\Pi^{e,m}=\sum_{n=1}^\infty\Pi_n^{e,m}$ with
\begin{align}
\Pi^{e,m}_n&=i^n\frac{2n+1}{n(n+1)}f_n^{e,m}(r)P_n^1(\cos\theta)\Phi^{e,m}(\varphi),\label{4}
\end{align}
in which $\Phi^e=\cos\varphi$, $\Phi^m=\sin\varphi$, and  $P_n^1$ denotes the associated Legendre function \cite{Abramowitz}. The radial functions $f_n^{e,m}$ satisfy the differential equations \begin{align}
&\mathcal{L}^{e,m}_n\left[rf_n^{e,m}\right]=0;\label{5}\\
&\mathcal{L}^{e,m}_n\equiv\frac{d^2}{dr^2}+\frac{d(\ln\delta^{e,m})}{dr}\frac{d}{dr}+k^2\epsilon-\frac{n(n+1)}{r^2},\label{5a}
\end{align}
with $\delta^e=\epsilon^{-1}$ and $\delta^m=1$, and the boundary conditions
\begin{align}
f_n^{e,m}&\text{ regular at }r=0;\label{6}\\
f_n^{e,m}(b)&=\left[j_n+a_n^{e,m}h_n^{(1)}\right]_{r=b};\label{7}\\
\delta^{e,m}\!\!\left.\frac{\left(rf_n^{e,m}\right)}{dr}\right|_{r=b}\!&=\frac{d}{dr}\left[rj_n+a_n^{e,m}rh_n^{(1)}\right]_{r=b}.\label{8}
\end{align}
The quantities $a_n^{e,m}$ in Eqs. (\ref{7})-(\ref{8}) are constants, and $j_n$ and $h_n^{(1)}$ denote the spherical Bessel and Hankel functions of the first kind \cite{Abramowitz} with argument $kr$, respectively. Eqs. (\ref{7})-(\ref{8}) ensure proper matching of the internal and external fields at the plasma boundary. In Eqs. (\ref{2})-(\ref{4}), each partial potential $\Pi^{e,m}_n$ with the corresponding electric field $\bm{E}^{e,m}_n$ induces a particular oscillation mode of the electrons in the plasma \cite{Bohren}, which has a current distribution $\bm{J}_n^{e,m}\propto\bm{E}^{e,m}_n$. The radiation emitted from the plasma by the current $\bm{J}^{e,m}_n$ has the form of $n$th-order electric (e) or magnetic (m) multipole radiation, with an amplitude proportional to $a_n^{e,m}$.
\subsection{Quasistatic limit \label{sec2b}}
In sections \ref{sec4} and \ref{sec5}, we will calculate ponderomotive forces for concrete examples of subwavelength plasmas, based on the fields formulated in section \ref{sec2a}. However, for $kb\ll1$ the electric field inside the plasma can be approximated \cite{Bohren} by the quasistatic field $\bm{E}_{qs}\exp\left(-i\omega t\right)$, where $\bm{E}_{qs}$ is the field that would be present if $\bm{E}_{ext}$ were replaced by the static field $E_0\bm{e}_x$. Here, we therefore briefly describe this quasistatic field as well, so that the corresponding ponderomotive forces can be compared to the forces based on the full-wave electric field of section \ref{sec2a}. We find that both approaches often agree very well, as expected, which makes the quasistatic description a useful way to quickly gain an impression of the fields and forces in a subwavelength plasma. However, we will also show that certain important physical effects are completely missing from the quasistatic description. One should therefore always be careful when using this approximation, as the full-wave approach is imperative to reveal all aspects of the interaction of the plasma with the applied wave.\\

The field $\bm{E}_{qs}$ is determined by the Maxwell equations $\nabla\cdot\left(\epsilon\bm{E}_{qs}\right)=0$ and $\nabla\times\bm{E}_{qs}=\bm{0}$. Substituting in the static Maxwell equations
\begin{align}
\bm{E}_{qs}=-E_0\nabla\phi\label{17}
\end{align}
results in a partial differential equation for $\phi$ that can be separated in spherical coordinates by writing $\phi=\psi(r)Y(\theta,\varphi)$. Solutions for the angular part are the spherical harmonics, of which only the particular harmonic $Y=\sin\theta\cos\varphi$ suits the symmetry of the problem. Accordingly,
\begin{align}
\phi=\psi(r)\sin\theta\cos\varphi,\label{17a}
\end{align}
where the radial function $\psi(r)$ is determined by the differential equation
\begin{align}
\left[\frac{d^2}{dr^2}+\left(\frac{2}{r}+\frac{1}{\epsilon}\frac{d\epsilon}{dr}\right)\frac{d}{dr}-\frac{2}{r^2}\right]\psi=0.\label{18}
\end{align}
The accompanying boundary conditions are that $\psi$ be regular at $r=0$, that both $\psi$ and $\epsilon d\psi/dr$ be continuous at $r=b$, and that $-E_0\nabla\phi\rightarrow E_0\bm{e}_x$ as $r\rightarrow\infty$. These conditions evaluate to
\begin{align}
\psi(0)\text{ regular at }r&=0;\label{19}\\
\left(\epsilon\frac{d\psi}{dr}+\frac{2\psi}{r}\right)_{r=b}&=-3.\label{21}
\end{align}
The quasistatic solution (\ref{17})-(\ref{21}) also follows directly from the more general results of the previous section by taking the appropriate limits. This is shown in Appendix \ref{secA}.
\subsection{Real and imaginary parts of $\bm{f_n^{e,m}}$ \label{sec2c}}
Although the quasistatic field Eqs. (\ref{17})-(\ref{21}) is generally a good approximation when $kb\ll1$, it lacks certain features that are essential to describe a number of physical effects. As we will show later, the latter include the presence of a nonzero radiation pressure on the plasma and a compressive ponderomotive force in case of a uniform density profile. The description of these effects requires the use of the full-wave solution of Section \ref{sec2a}. In particular, the boundary conditions Eq. (\ref{7})-(\ref{8}), and hence the functions $f_n^{e,m}$, are in general complex-valued. The presence of the nonzero imaginary parts of $f_n^{e,m}$ leads to phase shifts in the corresponding fields contributions, and these phase shifts give rise to the mentioned physical effects. To describe these effects adequately in the next sections, we derive here a new representation for the functions $f_n^{e,m}$ in which the real and imaginary parts are conveniently separated. Eliminating the constants $a_n^{e,m}$ from Eqs. (\ref{7})-(\ref{8}) gives, at $r=b$,
\begin{align}
\delta^{e,m}\frac{d\left(rf_n^{e,m}\right)}{dr}-\frac{d(rh_n^{(1)})}{dr}\frac{f_n^{e,m}}{h_n^{(1)}}&=rh_n^{(1)}\frac{d(j_n/h_n^{(1)})}{dr}.\label{22}
\end{align}
Replacing the Bessel functions in Eq. (\ref{22}) by their limiting value for small argument \cite{Abramowitz}, it is apparent that the imaginary part of $f_n^{e,m}$ is very small. This suggests to define auxiliary functions $g_n^{e,m}$ that, like $f_n^{e,m}$, are regular solutions of the differential equation
\begin{align}
\mathcal{L}^{e,m}_n\left[rg^{e,m}_n\right]=0,\label{22a}
\end{align}
but instead with a real-valued boundary condition that at $r=b$
\begin{align}
\hspace{-1.6mm}\left[\delta^{e,m}\frac{d}{dr}-\!\left(\frac{1}{b}+\frac{d}{dr}\ln\left|h_n^{(1)}\right|\right)\right]rg_n^{e,m}&=-\frac{y_n}{kb\raisebox{2pt}{\big|}h_n^{(1)}\raisebox{2pt}{\big|}^2}.\label{28}
\end{align}
Here, $y_n$ denotes the spherical Bessel function of the second kind \cite{Abramowitz} with argument $kr$. Eq. (\ref{28}) has been obtained by replacing $f_n^{e,m}\rightarrow g_n^{e,m}$ in Eq. (\ref{22}) and taking the real part of the equation assuming real $g_n^{e,m}$. By construction, solution of Eqs. (\ref{22a})-(\ref{28}) yields real-valued functions $g_n^{e,m}$ that approximate the real part of $f_n^{e,m}$ for small $kb$. The imaginary part of $f_n^{e,m}$ can be extracted from $g_n^{e,m}$ as follows. Since $f_n^{e,m}$ and $g_n^{e,m}$ satisfy the same differential equations but different boundary conditions,
\begin{align}
f_n^{e,m}=\gamma_n^{e,m}g_n^{e,m},\label{29}
\end{align}
where $\gamma_n^{e,m}$ are constants. To determine these constants, we substitute Eq. (\ref{29}) in Eq. (\ref{22}), simplify the result by using Eq. (\ref{28}), and solve for $\gamma_n^{e,m}$. This gives
\begin{align}
\gamma_n^{e,m}&=1+\left.\frac{\left(j_n-g_n^{e,m}\right)\left(g_n^{e,m}+iy_n\right)}{y_n^2+(g_n^{e,m})^2}\right|_{r=b}\label{30}\\
&=1+i\left.\frac{j_n-g_n^{e,m}}{y_n}\right|_{r=b}+O\left[(kb)^{4n+2}\right].\label{31}
\end{align}
Eqs. (\ref{29})-(\ref{31}) give the real and imaginary parts of $f_n^{e,m}$ separately.
\section{Forces \label{sec3}}
\subsection{Ponderomotive volume and surface forces \label{sec3a}}
Gradients in the electric field formulated in section \ref{sec2} give rise to a ponderomotive volume force density according to Eq. (\ref{1}). In addition to this well-known volume force, there can also exist a ponderomotive surface force density or pressure $\pi_p$ acting on the boundary of the plasma. The presence of a surface force is easily estabished from Eq. (\ref{1}). Suppose that at $r=b$ the plasma density changes discontinuously from a finite value to zero, such that the permittivity discontinuously increases to unity. Then, because of the boundary conditions that both the perpendicular component of $\epsilon\bm{E}$ and the tangential component of $\bm{E}$ be continuous at $r=b$, the squared magnitude $E^2$ in Eq. (\ref{1}) must be discontinuous and $\nabla E^2$ must behave like a delta function. This singular feature represents an infinitely large volume force density present in a shell with infinitesimally small volume, that is, a surface force density. To evaluate this surface force density, we consider the total, integrated ponderomotive force $\bm{F}$ acting on the plasma. The integration volume $V$ is chosen to be a sphere with radius $b^+\equiv\lim_{\Delta\downarrow0}(b+\Delta)$ concentric with the plasma. Then, $V$ is split in two contributions as
\begin{align}
\bm{F}=\int\hspace{-1mm}\bm{f}\,dV^-+\iint_{b^-}^{b^+}\hspace{-2mm}\bm{f}r^2drd\Omega,\label{35}
\end{align}
where $b^-\equiv\lim_{\Delta\downarrow0}(b-\Delta)$, the volume $V^-$ is a sphere with radius $b^-$, and $\int d\Omega$ denotes integration over the angular coordinates. In this way, the singularity in the ponderomotive force density is contained in the second integral of Eq. (\ref{35}), so that this term will give the surface contribution to $\bm{F}$, while the first integral represents the ordinary ponderomotive volume forces. Furthermore, $\bm{f}$ may be written as the time-average of the divergence of a tensor \cite{Shearer}:
\begin{align}
\bm{f}=\left\langle\nabla\cdot\left(\epsilon_0\epsilon\bm{E}\bm{E}+\frac{1}{\mu_0}\bm{B}\bm{B}-U\bm{\mathrm{I}}\right)\right\rangle\equiv\left\langle\nabla\cdot\bm{\mathsf{T}}\right\rangle,\label{36}
\end{align}
where $\bm{\mathrm{I}}$ is the identity tensor, $U=(\epsilon_0E^2+\mu_0^{-1}B^2)/2$, and $\mu_0$ is the vacuum permeability, and angular brackets denote time-averaging. Using Eq. (\ref{36}) and Gauss' theorem for tensors \cite{Morse}, the second integral of Eq. (\ref{35}) may be rewritten as
\begin{align}
\iint_{b^-}^{b^+}\hspace{-2mm}\bm{f}r^2drd\Omega=\left\langle\int\! d\bm{\Omega}^+\!\cdot\bm{\mathsf{T}}-\int\! d\bm{\Omega}^-\!\cdot\bm{\mathsf{T}}\right\rangle,\label{37}
\end{align}
where $\Omega^\pm$ are spherical surfaces at $r=b^\pm$ with outward normal. Writing out the tensors in Eq. (\ref{37}), and using the boundary conditions for the fields to express all field components in terms of those at $r=b^-$, gives
\begin{align}
\bm{F}&=\int\hspace{-1mm}\bm{f}\,dV^--\int\pi_p\,d\bm{\Omega}^-,\label{38}
\end{align}
in which $\pi_p=-\epsilon_0\left(\epsilon-1\right)^2E_r^2/4$. The quantity $\pi_p$ represents an additional ponderomotive pressure that acts on the surface of a plasma with an abrupt plasma boundary. This pressure is always negative, corresponding to a surface force density in the outward direction. A surface force similar to Eq. (\ref{38}) has been obtained earlier for the special case of a plane wave refracted by a plane plasma boundary \cite{Klima}.\\

To some extent, the surface force density found here may appear to be an artifact of the ponderomotive force expression Eq. (\ref{1}). After all, in the plasma context this expression has originally been derived from a perturbation expansion of the equation of motion of single electrons \cite{Boot,Gaponov}, and in that sense seems to be an approximate quantity. However, the force Eq. (\ref{1}) follows identically \cite{Shearer} from the tensor in Eq. (\ref{36}), which in turn follows strictly from the thermodynamics of continuous media \cite{Landau}. Moreover, we have checked that integration of the arguably more fundamental averaged Lorentz force density $\langle\rho\bm{E}+\bm{J}\times\bm{B}\rangle$ gives the same result Eq. (\ref{38}). Furthermore, momentum conservation requires that the total force Eq. (\ref{38}) on the plasma balances the rate of momentum loss from the radiation field. As we will show in the next section, this is only the case in presence of the surface force density. Therefore, Eq. (\ref{38}) is the best that can be done within a continuum model of the plasma medium. Of course, the validity of the latter must break down at some point near the plasma boundary, which is essentially where the Debye length becomes comparable to the scale length of the plasma. Adequate modeling of the behavior of particles near the very plasma edge should therefore be based on particle tracking simulations invoking the full-wave expansion Eq. (\ref{2})-(\ref{4}) of the fields. However, this is outside the scope of this paper.
\subsection{Evaluation of the forces \label{sec3b}}
In order to facilitate practical application of the derived analytical results, we summarize the previous sections by listing explicit expressions for the various forces used in the remainder of the paper. Substituting the electric field Eqs. (\ref{2})-(\ref{4}) in Eq. (\ref{1}), and performing all differentiations, gives the following spherical components of the ponderomotive force density:
\begin{align}
f_j&=\chi\epsilon_0 kE_0^2\sum_{n=1}^\infty\sum_{m=n}^\infty\bigg\{\re\left(i^{m-n}\gamma_n^{m*}\gamma_m^m\right)R^{j1}_{nm}S^{j1}_{nm}\nonumber\\[-4pt]
&\hspace{11mm}+\re\left(i^{m-n}\gamma_n^{e*}\gamma_m^e\right)\left[R^{j2}_{nm}S^{j2}_{nm}+R^{j3}_{nm}S^{j3}_{nm}\right]\bigg\}\nonumber\\[-4pt]
&-\chi\epsilon_0 kE_0^2\sum_{n=1}^\infty\sum_{m=1}^\infty\im\left(i^{m-n}\gamma_n^{e*}\gamma_m^m\right)R^{j4}_{nm}S^{j4}_{nm},\label{32}
\end{align}
where $j=r,\theta,\varphi$ and $\chi\equiv\epsilon-1=-\omega_p^2/\omega^2$. The functions $R=R(r)$ and $S=S(\theta,\varphi)$ are listed in Appendix \ref{secB}. Note that the magnitude of the various contributions to the force essentially depend on the phase of the factors $\gamma^{e,m}_n$, which makes the formulation of section \ref{sec2c} particularly convenient for force calculations. Evaluation of the total ponderomotive force Eq. (\ref{38}) requires integration of Eq. (\ref{32}) over the plasma volume. The angular integrations can be performed analytically, and most terms in Eq. (\ref{32}) integrate to zero. The cartesian $x$- and $y$-components of $\bm{F}$ vanish completely in the integration over $\varphi$. In the remaining $z$ component, only terms with particular combinations of $n$ and $m$ survive the integration over $\theta$, which is shown in Appendix \ref{secB}. The resulting total volume force is
\begin{align}
\hspace{-2mm}\int\hspace{-1mm}\bm{f}\,dV^-&=-\frac{\pi\epsilon_0E_0^2}{k^2}\,\bm{e}_z\sum_{n=1}^\infty\bigg[\im\left(\gamma^{m*}_n\gamma^m_{n+1}\right)Y^{1}_n\label{33}\\[-4pt]
&+\im\left(\gamma^{e*}_n\gamma^e_{n+1}\right)\left(Y^{2}_n+Y^{3}_n\right)+\im\left(\gamma^{e*}_n\gamma^m_n\right)Y^{4}_n\bigg],\nonumber
\end{align}
where the quantities $Y_n^{1,2,3,4}$ are one-dimensional integrals over $r=0$ to $b^-$ involving the functions $g^{e,m}_n$; these integrals are given in Eqs. (\ref{B20})-(\ref{B23}). From Eq. (\ref{33}) it is apparent that only modes in the combinations $(\bm{E}^e_n,\bm{E}^e_{n+1})$, $(\bm{E}^m_n,\bm{E}^m_{n+1})$, and $(\bm{E}^e_n,\bm{E}^m_n)$ give nonzero contributions to the total ponderomotive volume force. That is, these are the combinations that give rise to a force density with a preferred direction. The surface force in Eq. (\ref{38}) involves only the electric (transverse magnetic) modes $\bm{E}^e_n$ since these are the only ones having a nonzero radial electric field component $E_r$. Analogous to the volume force above, in the angular integrations of Eq. (\ref{38}) all terms in $E_r^2$ vanish except for products $E^{e*}_{n,r}E^e_{n+1,r}$, resulting in
\begin{align}
-\!\!\int\hspace{-1mm}\pi_p\,d\bm{\Omega}^-&=-\frac{\pi\epsilon_0E_0^2}{k^2}\,\bm{e}_z\left.\frac{(\epsilon-1)^2}{\epsilon^2}\right|_{r=b^-}\label{34}\\[-5pt]
&\hspace{-1cm}\times\sum_{n=1}^\infty n(n+1)(n+2)\im\left(\gamma_n^{e*}\gamma_{n+1}^e\right)\left.g_n^eg_{n+1}^e\right|_{r=b^-}.\nonumber
\end{align}
\section{Homogeneous plasma \label{sec4}}
In the previous two sections, the fields and force densities induced by an electromagnetic wave in a spherical plasma with arbitrary $\epsilon(r)$ were formulated. In the remainder of the paper, we apply the results to a number of practical density profiles. Here, we start with plasmas with uniform density, which is one of the few density profiles for which analytical expressions for the fields are available. This will enable us to validate the results of the previous sections. In addition, the limit of small radius allows for simple analytical expressions for both the ponderomotive force density and the total force. This yields some interesting new insights in the way radiation interacts with subwavelength objects.
\subsection{Fields \label{sec4a}}
We first verify the field expressions of section \ref{sec2}. For a uniform plasma density giving a constant relative permittivity $\epsilon_1$, Eq. (\ref{5}) reduces to the spherical Bessel differential equation, and the expressions of section \ref{sec2a} reduce to the well-known Mie results \cite{Bohren}. For the quasistatic case $kb\ll1$ of section \ref{sec2b}, Eq. (\ref{18}) reduces to the Euler differential equation, and it is found that $\psi=-3r/(\epsilon_1+2)$. This gives $\bm{E}_{qs}=3E_0\bm{e}_x/(\epsilon_1+2)$, which is the well-known constant electric field in a homogeneous material sphere placed in a uniform static field \cite{Panofsky}, or the Mie solution in the Rayleigh limit $kb\rightarrow0$ \cite{Bohren}.\\
Using the functions $g_n^{e,m}$ of section \ref{sec2c} to evaluate the fields yields
\begin{align}
g_n^{e,m}=A_n^{e,m}j_n(\sqrt{\epsilon_1}kr),\label{38a}
\end{align}
where the constants $A_n^{e,m}$ are obtained from the boundary condition Eq. (\ref{28}). Explicit expressions are given in Appendix \ref{secC}. It is also shown there that the functions $f_n^{e,m}=\gamma_n^{e,m}g_n^{e,m}$, from which the potentials Eq. (\ref{4}) are generated, are equal to
\begin{align}
f_n^e=\sqrt{\epsilon_1}d_nj_n(\sqrt{\epsilon_1}kr);\hphantom{==}f_n^m&=c_nj_n(\sqrt{\epsilon_1}kr),\label{39}
\end{align}
where $c_n$ and $d_n$ are the coefficients of the internal field of the Mie solution in the customary formulation \cite{Bohren}. Comparison of the field definitions Eqs. (\ref{2})-({\ref{4}) with those of the Mie solution \cite{Bohren} indeed confirms Eq. (\ref{39}). All results of section \ref{sec2} thus correctly reduce to the Mie solution in the special case of uniform permittivity.
\subsection{Ponderomotive compression \label{sec4b}}
For a homogeneous plasma, the ponderomotive force density Eq. (\ref{32}) is readily evaluated by substituting Eq. (\ref{38a}), using the results Eqs. (\ref{C1})-(\ref{C2}) for $A^{e,m}_n$ and $\gamma^{e,m}_n$. For the general case, this gives a series of elaborate expressions in terms of Bessel functions. A more manageable result is obtained in the small radius limit $kb\ll1$, where the first few terms of the power series expansions Eqs. (\ref{C5})-(\ref{C10}) for $A^{e,m}_n$ and $\gamma^{e,m}_n$ suffice. Using the latter in Eq. (\ref{32}) gives, after considerable but straightforward algebra, the following lowest-order $(x,y,z)$-components of the ponderomotive force density:
\begin{align}
\hspace{-0.8mm}\bm{f}&=\frac{-\chi_1^2\epsilon_0k^2E_0^2}{10\left(\epsilon_1+2\right)^2\left(2\epsilon_1+3\right)^2\left(3\epsilon_1+4\right)}\!\left(\begin{array}{c}ux\\vy\\wz\end{array}\right)\!+\dots;\label{39a}\\[2pt]
\hspace{-0.8mm}u&=458+807\epsilon_1+432\epsilon_1^2+43\epsilon_1^3-15\epsilon_1^4;\nonumber\\
\hspace{-0.8mm}v&=3\left(2\epsilon_1+3\right)^2\left(18+13\epsilon_1\right);\nonumber\\
\hspace{-0.8mm}w&=416+794\epsilon_1+469\epsilon_1^2+61\epsilon_1^3-15\epsilon_1^4,\nonumber
\end{align}
where $\chi_1=\epsilon_1-1$ and the dots represent terms of order $O(b^3,r^3)$. Interestingly, Eq. (\ref{39a}) shows that a ponderomotive force density is present in the plasma which scales linearly with position in all three $(x,y,z)$-directions. Figure \ref{fig1} shows the corresponding three 'spring constants' $df_x/dx,df_y/dy,df_z/dz$ as function of $\epsilon_1$. Remarkably, the magnitude of the force density is almost equal in all directions irrespective of $\epsilon_1$. This is despite the fact that the exciting electromagnetic wave is not at all spherically symmetric, but propagates in the $z$-direction and is polarized in the $x$-direction. Moreover, the sign of each force component is opposite to that of the corresponding coordinate. Eq. (\ref{39a}) thus represents an almost isotropic, compressive ponderomotive force. In Fig. \ref{fig1}, the force correctly vanishes for $\epsilon_1\rightarrow1$, that is, in the limit of an infinitely rarified plasma. We deliberately displayed only underdense plasmas to avoid the complication of plasma resonances. The latter necessitate a more detailed model of the permittivity including damping, which is outside the scope of this paper. However, since no assumptions about the particular form of $\epsilon_1$ have been made in deriving Eq. (\ref{39a}), this expression is valid as well for more detailed descriptions of the plasma medium.\\

Note that any compression is completely absent in the quasistatic description, which predicts a perfectly constant electric field in the plasma and hence a vanishing ponderomotive force. The full-wave description of section \ref{sec2a} is therefore essential to obtain Eq. (\ref{39a}). We also remark that Eq. (\ref{39a}) has some analogy with the magnetic pinch force familiar from stationary currents, which is due to the self-generated magnetic field. In the case of our small driven plasma, a representative magnitude of the current densities present in the plasma is that of the electric dipole mode, which is $\bm{J}^e_1\approx-3i\epsilon_0\chi_1\omega E_0\bm{e}_x/(\epsilon_1+2)$. According to the Biot-Savart law \cite{Panofsky}, a hypothetical spherical medium carrying a stationary current density $\bm{J}^e_1$ would generate a magnetic field equal to $\mu_0\bm{J}^e_1\times\bm{r}/3$. The resulting Lorentz force density would be directed toward the $x$-axis and would have a magnitude of $-3\epsilon_0\chi_1^2k^2E_0^2\sqrt{y^2+z^2}/(\epsilon_1+2)^2$. The similarity with Eq. (\ref{29}) is evident, both regarding the magnitude and the proportionality with position. The driven plasma we consider, of course, is more complex than this crude model because the currents are both time-varying and have more structure than $\bm{J}^e_1$. In addition, electric forces play an equally important role. For these reasons, the ponderomotive force turns out to be Eq. (\ref{39a}) rather than the force just described, that is, the force is approximately radially compressive rather than pinching toward a single axis.
\begin{figure}[h]
\setlength{\abovecaptionskip}{20pt}
\hspace{7mm}
\begin{overpic}[width=0.85\columnwidth,trim=0mm 0mm 0mm 0mm,clip=true,grid=false]{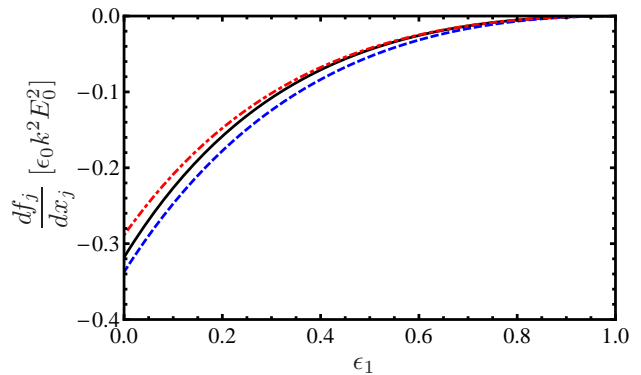}
\put(-6,35){\rotatebox{90}{\makebox(0,0){$\dfrac{df_j}{dx_j}\,\,[\epsilon_0k^2E_0^2]$}}}
\put(52,-3){\makebox(0,0){$\epsilon_1$}}
\end{overpic}
\caption{Cartesian $x$- (black solid), $y$- (blue dashed) and $z$- (red dash-dotted) components of the ponderomotive force Eq. (\ref{39a}), divided by the corresponding coordinate, as a function of the permittivity of the plasma.}
\label{fig1}
\end{figure}
\\

Ponderomotive compression by means of the force Eq. (\ref{39a}) seems interesting for technological applications such as confinement of spherical subwavelength plasmas. However, Eq. (\ref{39a}) is in fact the lowest-order correction to the ponderomotive force due to the quasistatic field, which coincidentally vanishes for the special case of a homogeneous plasma. For other than uniform density profiles, the ponderomotive force is dominated by the inhomogeneous quasistatic field, as we will show in the next section. Therefore the applicability of Eq. (\ref{39a}) to practical plasmas is limited. On the other hand, Eq. (\ref{39a}) is very relevant in scattering experiments where other media with a well-defined constant density, such as water droplets, are subjected to electromagnetic radiation \cite{Duft,Ward,Lettieri}. In addition, delicate physical processes that require contact-free observation of levitated droplets, such as surface vibrations \cite{Hill}, ice nucleation \cite{Stockel}, and crystallization of salts \cite{Wolf}, may be manipulated ponderomotively by application of an electromagnetic wave.
\subsection{Total ponderomotive force \label{sec4c}}
We next consider the total force on the plasma caused by the interaction with the incident wave. In scattering theory, the total force due to an incident wave is usually not formulated in terms of force densities, but rather is derived by calculating the rate at which momentum is carried away by the scattered radiation in the far field. This rate is identified with the total force on the body on account of momentum conservation \cite{[{For recent examples of such calculations, see }]Bekshaev,*Gomez}. In terms of the scattering coefficients $a^{e,m}_n$ in Eqs. (\ref{7})-(\ref{8}), the force reads \cite{Debye,Kerker}
\begin{align}
&\bm{F}=\frac{2\pi}{k^2}\frac{I}{c}\bm{e}_z\re\sum_{n=1}^\infty\left[\vphantom{\frac{2n(n+2)}{n+1}}\left(2n+1\right)\left(a^e_n+a^m_n\right)\right.\label{39b}\\
&\!\!-\left.\frac{2n(n+2)}{n+1}\left(a^{e*}_na^e_{n+1}+a^{m*}_na^m_{n+1}\right)-\frac{2(2n+1)}{n(n+1)}a^{e*}_na^m_n\right].\nonumber
\end{align}
In case of a small dielectric spherical scatterer with uniform permittivity $\epsilon_1$ and radius $b\ll k^{-1}$, Eq. (\ref{39b}) gives the following expansion \cite{Debye,Kerker}:
\begin{align}
\bm{F}=&\frac{8\pi k^4b^6}{3}\frac{\chi_1^2}{(\epsilon_1+2)^2}\frac{I}{c}\,\bm{e}_z\label{40}\\
&\times\left(1-\frac{120+34\epsilon_1-29\epsilon_1^2+\epsilon_1^3}{(\epsilon_1+2)(2\epsilon_1+3)}(kb)^2+\dots\right).\nonumber
\end{align}
Although certainly correct, this derivation of Eq. (\ref{40}) does not give any information about the distribution of the force over the scatterer. This is contrary to calculating $\bm{F}$ by integrating force densities such as in Eq. (\ref{38}), where one starts from the force distribution itself. In particular, it becomes clear that only part of the force is acting on the bulk, the remainder presenting itself in the form of a surface force. To our knowledge, such a direct analytical evaluation of the force on a scattering sphere from the local fields has never been given, although the force Eq. (\ref{40}) has been reproduced for special cases by numerically integrating the Maxwell stress tensor over the surface of the sphere \cite{Drobnik}, and by adding numerically calculated forces on a grid of dipoles representing the sphere \cite{Hoekstra}. Nevertheless, the force integration Eq. (\ref{38}) also correctly leads to Eq. (\ref{40}). Namely, substituting Eq. (\ref{38a}) together with the results (\ref{C5})-(\ref{C10}) in the force expressions Eqs. (\ref{33})-(\ref{34}), it is found that
\begin{align}
\int\hspace{-1mm}\bm{f}\,dV^-\!&=\frac{8\pi k^4b^6}{3}\frac{\chi_1^2}{(\epsilon_1+2)^2}\frac{I}{c}\,\bm{e}_z\label{41}\\
&\hspace{-7mm}\times\left(\frac{\epsilon_1+4}{2\epsilon_1+3}-\frac{Q_1}{210(\epsilon_1+2)(2\epsilon_1+3)^2}(kb)^2+\dots\right);\nonumber\\
-\!\!\int\hspace{-1mm}\pi_p\,d\bm{\Omega}^-\!&=\frac{8\pi k^4b^6}{3}\frac{\chi_1^2}{(\epsilon_1+2)^2}\frac{I}{c}\,\bm{e}_z\label{42}\\
&\hspace{-7mm}\times\left(\frac{\chi_1}{2\epsilon_1+3}-\frac{\chi_1Q_2}{70(\epsilon_1+2)(2\epsilon_1+3)^2}(kb)^2+\dots\right),\nonumber
\end{align}
with
\begin{align*}
Q_1&\equiv6720+3342\epsilon_1-1055\epsilon_1^2-215\epsilon_1^3+28\epsilon_1^4;\\
Q_2&\equiv560+78\epsilon_1-185\epsilon_1^2.
\end{align*}
Adding Eqs. (\ref{41})-(\ref{42}) reproduces the total ponderomotive force Eq. (\ref{40}) that was derived from momentum conservation. This confirms the validity of Eq. (\ref{38}).
\begin{figure}[h]
\setlength{\abovecaptionskip}{15pt}
\hspace{2mm}
\begin{overpic}[width=0.9\columnwidth,trim=0mm 0mm 0mm 0mm,clip=true,grid=false]{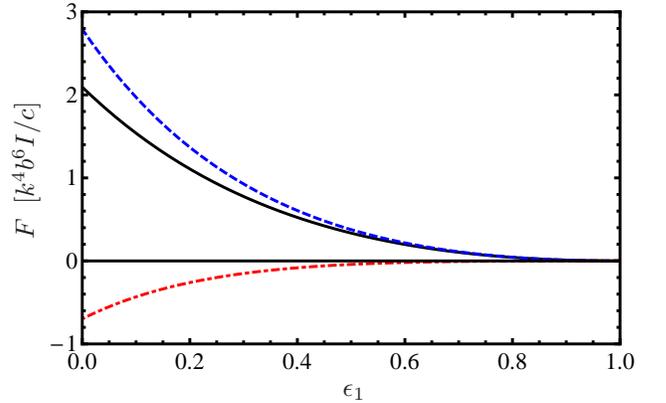}
\put(-4,35){\rotatebox{90}{\makebox(0,0){$F\,\,\,[k^4b^6I/c]$}}}
\put(52,-3){\makebox(0,0){$\epsilon_1$}}
\end{overpic}
\caption{Division of total ponderomotive force Eq. (\ref{38}) (black solid) into the volume contribution Eq. (\ref{41}) (blue dashed) and surface contribution Eq. (\ref{42}) (red dash-dotted) as a function of the permittivity of the plasma, for sufficiently small $kb$.}
\label{fig2}
\end{figure}
\\

In Eqs. (\ref{41})-(\ref{42}), the first terms in the large braces are dominant for small $kb$. Interestingly, the volume and surface contributions to the total force act in opposite directions, since $\chi_1$ is negative for plasmas. Furthermore, the division of the total ponderomotive force into the volume and surface contributions is dependent on $\epsilon_1$, which is shown in Fig. \ref{fig2}. As before, the forces correctly vanish in the limit $\epsilon_1\uparrow1$ of an infinitely rarified plasma. The ratio of the magnitude of the surface contribution to that of the volume contribution grows as $\epsilon_1$ drops, increasing to as much as $1/4$ for $\epsilon_1=0$. This shows the ponderomotive surface force derived in section \ref{sec3a} is not merely a small correction to the conventional volume ponderomotive force, but rather is an essential ingredient in a correct local description of the radiation pressure on subwavelength objects.\\

Finally, we note that we have only considered the limit $kb\ll1$ here. It would be interesting to show analytically the equality of Eq. (\ref{38}) with the general expression Eq. (\ref{39b}) for arbitrary $kb$. It is encouraging that the products of scattering coefficients in the second line of Eq. (\ref{39b}) represent the same combinations of modes that contribute to the integrated ponderomotive volume force Eq. (\ref{33}). On the other hand, the single coefficients in the first line of Eq. (\ref{39b}) do not have an analogue in Eq. (\ref{33}), which suggests that it is probably necessary to use certain special properties as well as recurrence relations for the Mie coefficients \cite{BohrenRecurrence}.
\section{Inhomogeneous plasmas \label{sec5}}
The homogeneous plasma considered above allowed us to validate the analytical results of sections \ref{sec2} and \ref{sec3}. In this section, we proceed to plasmas with radially varying density profiles. Lacking analytical solutions to the differential equations (\ref{22a}) that determine the fields, the results will be necessarily numerical. Experimentally, nanoplasmas that are field-ionized by laser pulses usually exhibit a natural density profile in which $dn_e/dr<0$ everywhere. In contrast, ultracold plasmas may be created with any desired density profile by photo-ionizing an atomic cloud using imaging techniques \cite{McCulloch}. In particular, 'inverted' profiles in which $dn_e/dr>0$ in some range of $r$ are possible. Such an inverted profile also results naturally when using sufficiently dense atomic clouds, that in their central region are optically thick for the excitation laser involved in the ionization scheme.
\subsection{Ponderomotive force distribution \label{sec5a}}
We have calculated the distribution of the ponderomotive force density for several density profiles by numerically solving the boundary value problem Eqs. (\ref{22a})-(\ref{28}) for the first few modes, and subsequently evaluating Eq. (\ref{32}) truncated at $n\leq3,m\leq3$. We have concentrated on subwavelength plasmas with $kb\sim0.1$, so that the truncated series proved to be sufficient to approximate the exact force density accurately. A shooting method was used to solve Eqs. (\ref{22a})-(\ref{28}), in which the numerical stability was improved by switching variables from $g^{e,m}_n$ to $x^{e,m}_n=g^{e,m}_n/(kr)^n$, and avoiding the singular point at $r=0$ by imposing the condition $dx^{e,m}_n/dr=0$ at a finite radius $r=r_0\ll b$. Decreasing $r_0$ to $0.01b$ yielded sufficiently converged results.\\

In order to test our numerical code, we have calculated the force density in a homogeneous plasma with a smoothed edge according to the density profile $n_e(r)=\left\{1-\tanh\left[\alpha\left(r/b-0.9\right)\right]\right\}n_0/2\equiv n_1(r)$. Choosing $kb=0.11$, this profile represents a plasma with density $n_0$, which at radius $0.1k^{-1}$ drops smoothly to zero within a small distance of about $4/(\alpha k)$. In Fig. \ref{fig3}(a), this profile is shown as the red dash-dotted line, with the corresponding vertical axis on the right of the figure. Since the calculational domain extends to $r=b$, the force density is thus evaluated up to radii outside the plasma, rather than up to an arbitrary point somewhere in the edge region $r\approx0.1k^{-1}$. The benefit of this approach over using the simpler discontinuous profile $n_e(r)=n_0\Theta\left(r-b\right)$ is that it is possible to study the volume force density in the edge region, which must tend to the surface force density in Eq. (\ref{38}) as $\alpha\rightarrow\infty$. In the calculations, the plasma density $n_0$ was chosen such that $\epsilon_1\equiv1-n_0e^2/(\epsilon_0m_e\omega^2)=0.19$. In Fig. \ref{fig3}, the radial component of the volume force density Eq. (\ref{32}) along the positive $x$-axis is shown by the black solid line, for $\alpha=200$. We have deliberately chosen to present the $x$-direction, which is the polarization axis of the incident wave, because in this direction there is a strong radial electric field component $E_r$. Hence the surface force density defined in Eq. (\ref{38}) is clearly exhibited, in contrast to some other directions such as the $y$-axis in which the surface force vanishes.\\
It was derived in section \ref{sec4b} that, in the bulk of the plasma, the ponderomotive force density should be compressive and proportional to $r$, and given by Eq. (\ref{39a}). The latter result is indicated in Fig. \ref{fig3}(a) by the blue dots. The numerical data closely follow the analytical result, which validates our numerical code. The black dashed line in Fig. \ref{fig3}(a) show the force density according to the quasistatic electric field, which is determined by Eq. (\ref{17})-(\ref{21}). As expected, the linear ponderomotive force is absent from the quasistatic description because the latter predicts a uniform electric field in the plasma bulk.\\

In Fig. \ref{fig3}(a) near the plasma edge at $kr=0.1$, the ponderomotive force density exhibits a steep positive peak. Fig. \ref{fig3}(b) is a close-up of the edge region, showing that this peak is positioned just on the inner side of the plasma edge. Note that the quasistatic field is also sufficient to correctly describe this feature, since the solid and dashed curves overlap perfectly. We found that for increasing values of $\alpha$, the peak becomes ever higher and narrower, but the energy density defined by the surface area below the peak $u_0=\int\!f_rdr$ stays approximately constant. This suggests that the peak will tend to the surface force density in Eq. (\ref{38}) as $\alpha\rightarrow\infty$. The surface area $u_1$ represented by the latter is obtained by writing the surface force density $\pi_p\bm{e}_r$ at position $(x,y,z)=(b,0,0)$ as the volume force density $\bm{f}_p\equiv\pi_p\delta\left(r-b\right)\bm{e}_r$. Integrating $f_{p,r}$, and using the quasistatic approximation $E_r\approx3E_0/(\epsilon_1+2)$, gives for the present case
\begin{align}
u_1=\int_{b^-}^{b^+}\hspace{-2mm}f_{p,r}dr=\frac{9\epsilon_0\chi_1^2E_0^2}{4(\epsilon_1+2)^2}=0.308\epsilon_0E_0^2.\label{43}
\end{align}
Numerical integration of a spline interpolation of the peak in Fig. \ref{fig3}(b) gives $u_0=0.307\epsilon_0E_0^2$, in excellent agreement with Eq. (\ref{43}). This confirms that the peaked volume force is the analogue of the surface force present in the limit of a discontinuous plasma boundary.\\

The outward ponderomotive force in the plasma edge region is reminiscent of a similar force that is found in case of a one-dimensional stratified plasma layer irradiated by a plane wave \cite{Hora}. However, the latter force is usually obtained by resorting to the WKB approximation to find the electric field, which is valid only when the plasma scale length is much larger than the wavelength. This is clearly not applicable for the subwavelength plasmas considered here. Furthermore, in the one-dimensional large scale length case, the force is proportional to $-\nabla n_e$ \cite{Hora}. This is not found in our case either, as evidenced by the fact that the peak in Fig. \ref{fig3}(b) does not coincide with the inflection point of the density at $kr=0.1$.
\begin{figure}[h]
\setlength{\abovecaptionskip}{15pt}
\hspace{1mm}
\begin{overpic}[width=0.9\columnwidth,trim=7mm 5mm 5mm 3mm,clip=true,grid=false]{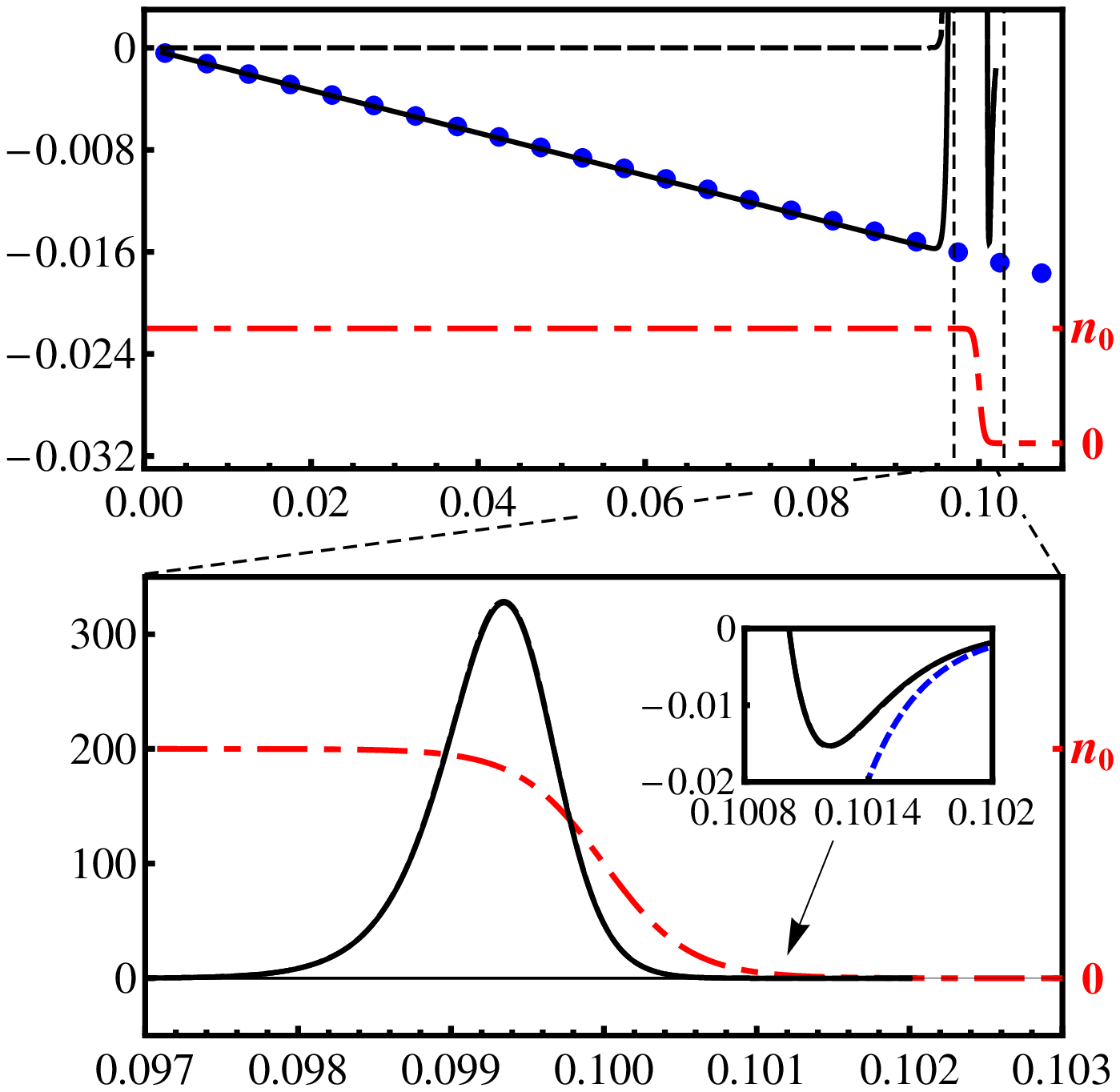}
\put(-3,94){\makebox(0,0){$(a)$}}
\put(-3,43){\makebox(0,0){$(b)$}}
\put(-3,75){\rotatebox{90}{\makebox(0,0){$f_r\,\,\,[\epsilon_0kE_0^2]$}}}
\put(2,25){\rotatebox{90}{\makebox(0,0){$f_r\,\,\,[\epsilon_0kE_0^2]$}}}
\put(102,21.5){\rotatebox{90}{\makebox(0,0){\color{red} $n_e$}}}
\put(102,63){\rotatebox{90}{\makebox(0,0){\color{red} $n_e$}}}
\put(54,-3){\makebox(0,0){$kr$}}
\end{overpic}
\caption{Radial component of the ponderomotive force density along the line $y=z=0$, for a homogeneous plasma with smoothed edge. (a) Force density in the plasma bulk according to numerical evaluation of Eq. (\ref{32}) assuming the full field (black solid), numerical evaluation of Eq. (\ref{1}) assuming the quasistatic approximation (black dashed), and the analytical result Eq. (\ref{39a}) (blue dots). For orientation, the plasma density (red dot-dashed line) is shown together with the results. (b) Close-up of the edge region indicated by the vertical dashed lines in the upper panel. The dashed and solid curves overlap. The inset is a close-up of the horizontal axis, showing also the force Eq. (\ref{1}) assuming the field of an equivalent electric dipole (blue dashed).}
\label{fig3}
\end{figure}
\\

In Fig. \ref{fig3}(b), at the right side of the peak the force has a small overshoot to negative values, which is shown in the inset. The overshoot is visible as well in Fig. \ref{fig3}(a). The overshoot is caused by the inhomogeneous electric field outside the plasma, which is approximately that of an oscillating electric dipole \cite{Panofsky}. Since the plasma density has not yet completely vanished around $kr=0.1014$, the electric field gradient present there leads to a small but finite negative ponderomotive force density. The blue dashed line in the inset of Fig. \ref{fig3}(b) shows the force density Eq. (\ref{1}) assuming the mentioned dipole field. The numerical result indeed approaches this line.\\

Figure \ref{fig4} shows the ponderomotive force density for the plasma profiles $n_e(r)=\left\{3\pm\left[1-20(kr)^2\right]\right\}n_1(r)/4$, where $n_1(r)$ was defined at the beginning of this section, again evaluated along the positive $x$-axis. The profile with a plus (minus) sign represents a plasma with a quadratic bulge (dip) of the density in the central region, but with the same smoothed edge as in Fig. \ref{fig3}. The most important difference with respect to the flat profile discussed above is that the force density in the plasma bulk is significantly larger than the linearly varying force density shown in Fig. \ref{fig3}(a). This is because already in the quasistatic approximation, the electric field strength for the profiles of Fig. \ref{fig4} is inhomogeneous, whereas in the plasma of Fig. \ref{fig3} it is constant. Therefore, the ponderomotive force depicted in Fig. \ref{fig3} consists of merely small corrections to the vanishing contribution of the quasistatic field, whereas in Fig. \ref{fig4}(a) and \ref{fig4}(c) the force is completely dominated by the nonzero gradient of the quasistatic field itself. This is confirmed by the fact that the quasistatic and exact results in Fig. \ref{fig4} overlap perfectly.\\

Interestingly, the direction of the ponderomotive force in Figs. \ref{fig4}(a) and \ref{fig4}(c) depends on the type of plasma profile: for natural profiles with $dn_e/dr<0$, the force is directed outward; for inverted profiles with $dn_e/dr>0$, the force is directed toward the plasma center. This suggests that it is possible, at least regarding the plasma bulk, to tailor the ponderomotive force distribution by choosing a suitable initial density profile. For instance, it may be possible to devise a plasma in which ponderomotive forces balance hydrodynamic forces locally, which would mean that the plasma is stabilized rather than disturbed by application of an electromagnetic wave. However, the freedom to manipulate the ponderomotive force density is much more restricted in the edge region. Regardless of the type of density profile, at the plasma boundary the steep gradient in the plasma density invariably leads to the strongly peaked and outward ponderomotive force density found before, as is illustrated by Figs. \ref{fig4}(b) and \ref{fig4}(d). Obviously, this outward force is unfavorable for the stability of the plasma as it will tend to push electrons outwards.
\begin{figure}[h]
\setlength{\abovecaptionskip}{15pt}
\hspace{7.5mm}
\begin{overpic}[width=0.82\columnwidth,trim=12mm 2mm 6mm 3mm,clip=true,grid=false]{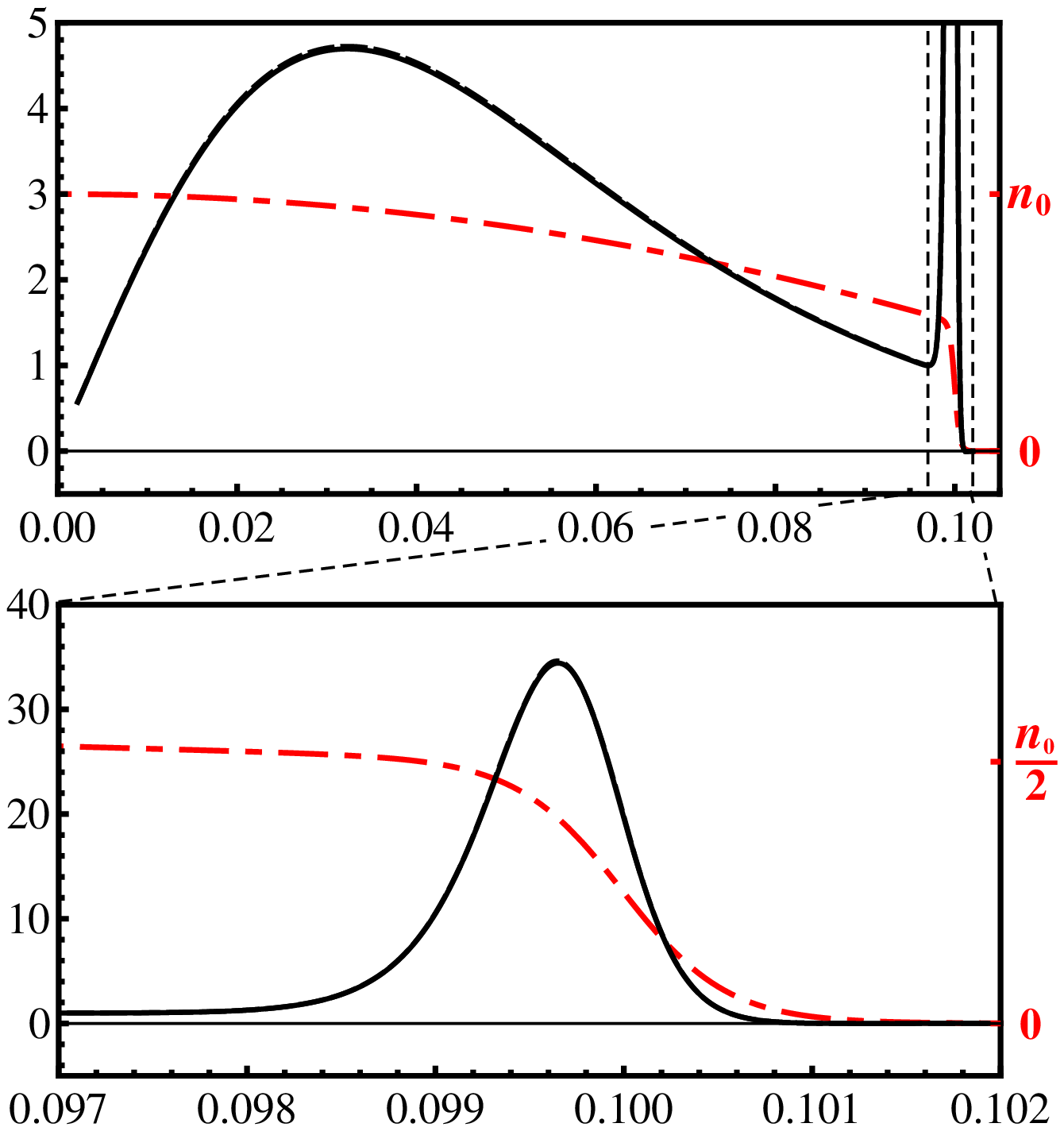}
\put(-12,94){\makebox(0,0){$(a)$}}
\put(-12,45){\makebox(0,0){$(b)$}}
\put(-7,75){\rotatebox{90}{\makebox(0,0){$f_r\,\,\,[\epsilon_0kE_0^2]$}}}
\put(-7,25){\rotatebox{90}{\makebox(0,0){$f_r\,\,\,[\epsilon_0kE_0^2]$}}}
\put(92,22){\rotatebox{90}{\makebox(0,0){\color{red} $n_e$}}}
\put(92,72){\rotatebox{90}{\makebox(0,0){\color{red} $n_e$}}}
\end{overpic}\\
\hspace{5mm}
\begin{overpic}[width=0.85\columnwidth,trim=5mm 3mm 4mm 1mm,clip=true,grid=false]{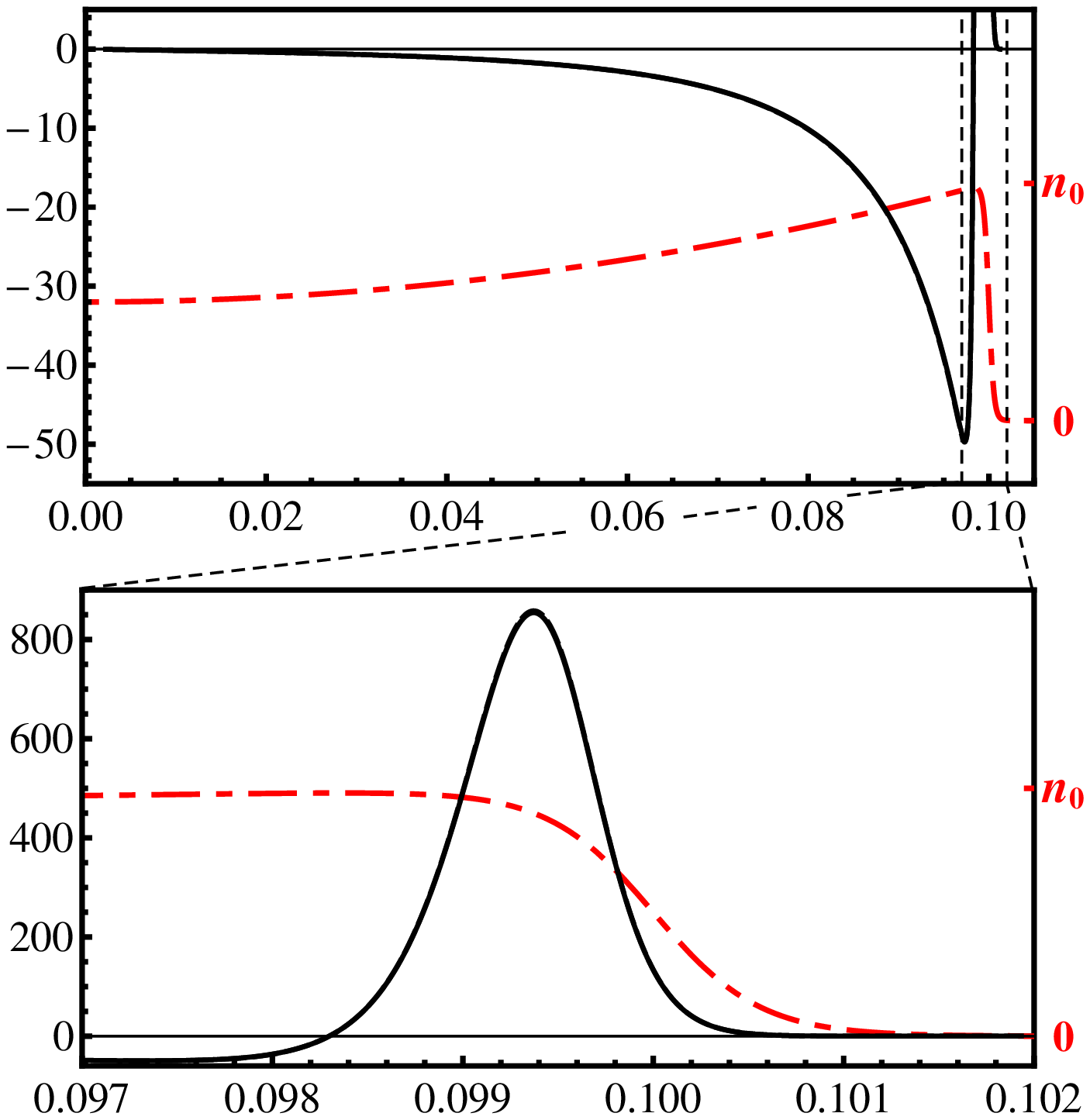}
\put(-9,94){\makebox(0,0){$(c)$}}
\put(-9,45){\makebox(0,0){$(d)$}}
\put(-4,75){\rotatebox{90}{\makebox(0,0){$f_r\,\,\,[\epsilon_0kE_0^2]$}}}
\put(-4,25){\rotatebox{90}{\makebox(0,0){$f_r\,\,\,[\epsilon_0kE_0^2]$}}}
\put(95,20){\rotatebox{90}{\makebox(0,0){\color{red} $n_e$}}}
\put(95,70){\rotatebox{90}{\makebox(0,0){\color{red} $n_e$}}}
\put(48,-1){\makebox(0,0){$kr$}}
\end{overpic}
\caption{Radial component of the ponderomotive force density along the line $y=z=0$, for plasma density profiles with a smoothed edge and a quadratic bulge (a),(b) and dip (c),(d). For further details, see Fig. \ref{fig3}.}
\label{fig4}
\end{figure}
\subsection{Total ponderomotive force \label{sec5b}}
We have calculated the total ponderomotive force acting on the plasmas considered in the previous section, by numerically evaluating the volume force integration Eq. (\ref{33}) truncated at $n\leq3$. The resulting forces are shown in Fig. \ref{fig5} as a function of the permittivity $\epsilon_1$. Crosses represent the data according to Eq. (\ref{33}). As a check, the forces have been calculated alternatively in terms of the scattered radiation, by numerically evaluating the scattering coefficients with Eq. (\ref{7}), and substituting these coefficients in Eq. (\ref{39b}). The resulting forces are shown in Fig. (\ref{fig5}) as open squares. Evidently, both methods agree very well, confirming the validity of Eq. (\ref{33}) for arbitrary density profiles.\\

For a given value of $\epsilon_1$, the total force on the plasma with a quadratic dip (D) is systematically smaller than that on the homogeneous plasma (H) with the same radius, and the force on the plasma with a quadratic bulge (B) is still smaller. This is easily explained in terms of the radiation scattered from the incident wave by the three plasmas. At the chosen plasma size $kb=0.11$, the electrons in the plasma move more or less coherently, so that the scattered radiation is predominantly electric dipole radiation with the radiated power proportional to the number $N$ of electrons squared. By conservation of momentum, the momentum lost from the incident wave and therefore the resulting total force on the plasma are proportional to $N^2$ as well. For the three plasmas considered in Fig. \ref{fig5}, equal $\epsilon_1$ implies equal densities $n_0$, resulting in squared numbers of electrons in the ratios $N^2_H:N^2_D:N^2_B=1:0.64:0.49$. These ratios roughly fit the relative heights of the curves in Fig. \ref{fig5}. However, the coherent model just given is not precise, first because both higher order multipole moments and directional asymmetry in the scattered radiation have been neglected, and second because profile dependent resonant behavior for $\epsilon_1$ near 0 has been disregarded. Nevertheless, we have numerically confirmed that the relative amplitudes of the total force on the three considered plasmas indeed tend to $N^2_H:N^2_D:N^2_B$ in the limits $kb\rightarrow0$ and $\epsilon_1\rightarrow1$ where the coherent model becomes exact.
\begin{figure}[h]
\setlength{\abovecaptionskip}{15pt}
\hspace{1mm}
\begin{overpic}[width=0.9\columnwidth,trim=0mm 0mm 0mm 0mm,clip=true,grid=false]{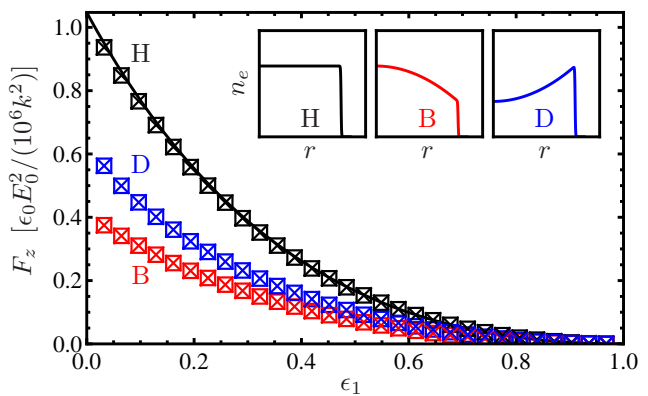}
\put(15,55){\makebox(0,0){H}}
\put(15,36){\makebox(0,0){\color{blue} D}}
\put(15,17){\makebox(0,0){\color{red} B}}
\put(44,44){\makebox(0,0){H}}
\put(64,44){\makebox(0,0){\color{red} B}}
\put(84,44){\makebox(0,0){\color{blue} D}}
\put(32,50){\rotatebox{90}{\makebox(0,0){$n_e$}}}
\put(44,38){\makebox(0,0){$r$}}
\put(64,38){\makebox(0,0){$r$}}
\put(84,38){\makebox(0,0){$r$}}
\put(-5,35){\rotatebox{90}{\makebox(0,0){$F_z\,\,\,[\epsilon_0E_0^2/(10^6k^2)]$}}}
\put(51,-2){\makebox(0,0){$\epsilon_1$}}
\end{overpic}\\
\caption{Total ponderomotive force as a function of the permittivity at the density $n_0$, according to scattering theory (numerical evaluation of Eq. (\ref{39b}), open squares), integration of the volume force Eq. (\ref{33}) (crosses), and Eq. (\ref{39b}) using the well-known Mie-coefficients (black solid line). Results are shown for a uniform profile (black, 'H') and profiles with a quadratic bulge (red, 'B') and dip (blue, 'D'); these profiles were defined in section \ref{sec5a} and have been sketched in the insets.}
\label{fig5}
\end{figure}
\section{Acceleration of ultracold plasmas \label{sec6}}
In the previous sections, we have carefully examined both the distribution of ponderomotive force in an electromagnetically driven subwavelength plasma, and the total resultant force derived from it by volume integration. In summary, it was found that in the plasma bulk the ponderomotive force is directed radially inwards for inverted density profiles, that a strongly localized outward force dominates near the very edge of the plasma, and that the total force on the plasma is approximately proportional to $N^2$. We are now in the position to assess the feasibility of practical acceleration of subwavelength plasmas based on the total ponderomotive force. This concept was put forward in the past by Veksler \cite{Veksler} and reviewed by Motz and Watson \cite{Motz}. The original formulation \cite{Veksler} of the acceleration mechanism was that subwavelength plasmas should scatter incident radiation at an energy rate of $N^2$ times the single electron value $\sigma_TI$, where $\sigma_T=e^4/(6\pi\epsilon_0^2m_e^2c^4)$ is the Thomson cross section. By conservation of momentum, this leads to a rate of momentum transfer to (or accelerating force on) the plasma of $N^2\sigma_TI/c$. Indeed, the total force Eq. (\ref{39b}) derived from the scattered radiation reduces to this force in the appropriate limits \cite{Motz}. What we have shown in this paper is that this force is equivalent to the integrated ponderomotive force in the plasma.\\

Acceleration experiments in the 1960s based on the above scheme have produced ions with keV energies \cite{Vekslerexp,Kononov}. However, static magnetic fields were necessary to confine the plasma in the transverse direction, and the exact acceleration mechanism was not very well understood \cite{Motz}. Moreover, the very large energy spread of the ions showed that the plasma was not accelerated as a compact bunch but rather completely dispersed over the length of the accelerator. These experiments were therefore discontinued in favor of more promising acceleration schemes. The reason why the radiative method can at the present time be more viable is the current availability of ultracold plasmas. Because the electron temperature of these plasmas is extremely low ($\sim10$ K), hydrodynamic forces are very small, so that any violent plasma expansion is absent. Moreover, as mentioned before, the density distribution of ultracold plasmas can easily be tailored to an inverted profile, either by means of imaging techniques or by using optically thick atomic clouds. As we have shown, the bulk ponderomotive force is compressive for inverted profiles, which could further reduce the plasma expansion.\\

Let us consider the velocities attainable by radiative acceleration. For this purpose, it is important to realize that in practice the plasma is not a rigid object, but will in general expand, so that not only $b$, but also the density and hence $\epsilon$ will vary with time. The number of particles $N$, on the other hand, remains approximately fixed. The accelerating total force will therefore depend on $b$ both directly through the coherence properties of the plasma and indirectly through its dependence on $\epsilon(b)$. Figure \ref{fig6} shows this dependency for three different $N$, assuming a driving frequency of $\omega/2\pi=1.3$ GHz (standard L-band microwaves) and a uniform density profile for which $\epsilon(b)=\epsilon_1=1-3N/(4\pi\epsilon_0m_e\omega^2b^3)$. Immediately apparent is the plateau in the force at $F_z/N^2=\sigma_TI/c$, indicated by the horizontal dashed line, which corresponds to the force proposed by Vesksler \cite{Veksler}. At the high $kb$ side, the force decreases rapidly once $kb\gtrsim1$ because the plasma electrons do no longer scatter incident radiation coherently at such larger plasma sizes. As this effect is a geometrical one, it is not dependent on the number of particles. At the low $kb$ side, each curve in Fig. \ref{fig6} strongly increases around the plasma radius $b_m$ at which the Mie resonance $\epsilon_1=-2$ occurs. This is where the driving frequency matches the eigenfrequency of oscillations of the whole electron cloud of the plasma in the field of the ion cloud \cite{Fennel}. Since $\epsilon_1$ depends on $N$, the radius $b_m$ is different for the three cases in Fig. \ref{fig6}, indicated by the vertical dashed lines. We have also calculated the total force for the other density profiles considered in this paper. This gives practically the same results on the scales of Fig. \ref{fig6}, although minor differences are found close to $b_m$ due to different resonance properties, and for $kb\gtrsim1$ due to different coherence properties. However, the plateau in the force is exactly the same, in accordance with the observation in section \ref{sec5b} that $F_z\propto N^2$ for all profiles if $\epsilon_1$ is close to unity, that is, away from the Mie resonance.
\begin{figure}[h]
\setlength{\abovecaptionskip}{15pt}
\hspace{1mm}
\begin{overpic}[width=0.895\columnwidth,trim=0mm 0mm 0mm 0mm,clip=true,grid=false]{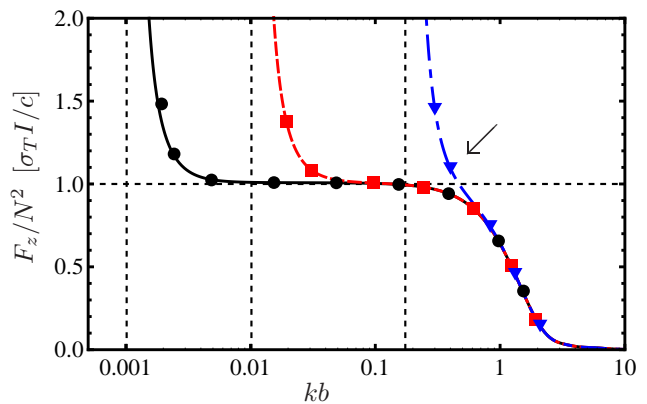}
\put(-5,35){\rotatebox{90}{\makebox(0,0){$F_z/N^2\,\,\,[\sigma_TI/c]$}}}
\put(45,-2){\makebox(0,0){$kb$}}
\put(71,39){\line(1,1){5}}
\put(71,39){\line(1,0){2}}
\put(71,39){\line(0,1){2}}
\end{overpic}
\caption{Total ponderomotive force as a function of plasma radius, when the particle number is fixed at $N=10^4$ (black solid line and dots), $N=10^7$ (red dashed line and squares), and $N=5\times10^{10}$ (blue dash-dotted line and triangles) according to Eq. (\ref{39b}) using the well-known Mie-coefficients (lines) and integration of the volume force Eq. (\ref{33}) (symbols). A uniform profile and a driving frequency of $\omega/2\pi=1.3$ GHz have been assumed. The horizontal dashed line represents the force according to coherently enhanced Thomson scattering; the vertical dashed lines indicate the radius at which Mie resonance occurs. The datapoint indicated by the arrow is discussed in the main text.}
\label{fig6}
\end{figure}
\\

Now, for acceleration purposes the plasma size should presumably be in the 'plateau range' of Fig. \ref{fig6}, in order to both have a significant acceleration and at the same time avoid plasma resonances. Experimentally, the latter invariably lead to significant electron loss and heating in both nanoplasma \cite{Fennel} and ultracold plasma \cite{Kulin,Twedt} experiments, and should therefore be avoided despite the greatly enhanced accelerating force. Secondly, the number of particles $N$ should be chosen as large as possible to maximize $F_z$. However, for too large $N$ the plateau range disappears as the resonance radius $b_m$ grows larger than $k^{-1}$. The dash-dotted curve in Fig. \ref{fig6} represents about the largest $N$ that allows for a plasma that is both coherent and non-resonant at the chosen driving frequency of 1.3 GHz. Incidentally, the corresponding value $N=5\times10^{10}$ is also one of the largest numbers of atoms that have actually been magneto-optically cooled and trapped \cite{Ketterle}. In that experiment, the atomic cloud consisted of sodium. Let us estimate what energies may be attained when this particular cloud is ionized and accelerated by 1.3 GHz microwave radiation. A suitable plasma radius, indicated in Fig. \ref{fig6} by the data point with an arrow, would be $1.5$ cm (even larger atomic clouds with sizes up to a few centimeters have been successfully produced \cite{Schoser}). Existing L-band klystrons \cite{Chin} can produce microwave pulses with length $\tau=1.5$ ms at a power exceeding 10 MW. At this power and with diffraction-limited focusing, the intensity is about $I=35$ kW/cm$^2$. The resulting electron oscillation amplitude is still much smaller than the plasma radius, so that the plasma should still behave as a dielectric as has been assumed in this paper. Assuming that $kb$ remains smaller than unity throughout the microwave pulse, the momentum transferred to the plasma is $p=\tau N^2\sigma_TI/c$. The corresponding kinetic energy per ion is $U=(p/N)^2/(2m_i)=2.7$ keV, where $m_i=3.8\times10^{-26}$ kg is the atomic mass of sodium. Thus the energies reachable by radiative acceleration are quite substantial.\\

Although keV energies are nowhere near those attained in conventional accelerators, it should be emphasized that an accelerated ultracold plasma is still an object with very special properties. First of all, it is an accelerated neutral beam, whereas other acceleration methods involve charged beams. An exception to some extent is acceleration of partially neutralized ion beams from laser-irradiated foils \cite{Henig}. However, in the latter method beam properties such as the energy spread ($\sim10\%$) are still poor. An accelerated ultracold plasma, on the other hand, may have remarkable beam quality. For ions at nonrelativistic energies, such as in the field of focused ion beams \cite{Orloff}, beam quality is usually expressed \cite{Luiten} in terms of the reduced brightness $B_r=eI_\text{peak}/(2\pi^2\xi^2m_ic^2)$, where $I_\text{peak}$ is the peak ion current and  $\xi=b\sqrt{k_BT_i/(m_ic^2)}/2$ is the transverse thermal emittance with $T_i$ the ion temperature. Present state-of-the-art ion beams, produced using liquid-metal ion sources \cite{Orloff}, have a brightness up to $B_r=10^6$ A/m$^2$srV. In case of our ultracold plasma, the temperature of the ion component usually equilibrates to a few Kelvin \cite{Killian}, resulting in an emittance of $\xi<1$ nm. The peak ion current is $I_\text{peak}=\pi b^2en_ev=0.06$ A, where $v=p/(Nm_i)$ is the velocity of the plasma, yielding a brightness of $B_r>10^5$ A/m$^2$srV. The brightness of the ions of an accelerated ultracold plasma may thus be comparable to that of existing high-performance ion sources, but with the important difference that an ultracold plasma is a neutralized beam.\\

The above estimates being encouraging, it is important to realize that they are based on the assumption that the plasma stays coherent throughout the ms microwave pulse, that is, that $kb\lesssim1$. However, one may expect that the low but finite electron temperature of the plasma leads to some plasma expansion due to the hydrodynamic pressure gradient $\nabla n_ek_BT_e$. On the other hand, for inverted plasma density profiles in which $dn_e/dr>0$, this gradient can be directed inwards, leading to compression rather than expansion. Moreover, as mentioned before, the ponderomotive force is directed inwards as well for inverted profiles, giving an additional compressive action. To fully assess the time-dependence of the plasma size, therefore, one should study the evolution of the density profile under influence of the self-consistent hydrodynamic and ponderomotive forces. Such an analysis is outside the scope of this paper. We do note that the characteristic hydrodynamic expansion rate of usual undriven, Gaussian ultracold plasmas is $db/dt\sim\sqrt{m_i/(k_BT_e)}$ \cite{Killian}. If the plasma considered above would expand at this rate with $T_e=10$ K, it would still take some 0.4 ms before the plasma grows larger than $kb=1$. The interaction time $\tau$ assumed above is of the same order of magnitude and therefore seems reasonable.\\

Another assumption made above is that the plasma does not appreciably heat up due to the microwave interaction. In absence of plasma resonances, the most important heating mechanism \cite{Smorenburg} is inverse Bremsstrahlung due to electron-ion collisions. In the strong-field regime $e^2E_0^2/(4m_e\omega^2)\gg k_BT_e$ under consideration here, the electron-ion collision rate is  $\nu_{ei}\sim n_eem_e\omega^3/(\pi^2\epsilon_0^2E_0^3)$ \cite{Silin}, and the resulting heating rate per electron is $P_{ei}=\nu_{ei}e^2E_0^2/(2m_e\omega^2)$. In the example above, $\nu_{ei}\sim3$ s$^{-1}$ only, giving $P_{ei}=10^{-19}$ W. This corresponds to a temperature increase of only $P_{ei}/k_B=8$ K/ms. The plasma should therefore indeed remain ultracold during the acceleration process.\\

Finally, we should mention the strongly peaked outward ponderomotive near the edge of the plasma, which is of course disadvantageous for the stability of the plasma. Initially, the electrons in the edge region will probably be expelled from the plasma by this force. However, very soon, after a sufficient number $N_1$ of electrons has escaped, the resulting charging of the plasma will prevent any further electron loss. This happens as soon as the Coulomb potential $U_C=N_1e^2/(4\pi\epsilon_0b)$ of the plasma is larger than the kinetic energy $U_1$ that can be supplied to an electron by the ponderomotive force peak. The latter equals $U_1=u_1/n_e$, where $u_1$ is given by Eq. (\ref{43}) for a homogeneous plasma. For the plasma considered in this section, the condition $U_C=U_1$ gives $N_1/N=0.3\%$ only. Electron loss due to the ponderomotive force peak at the plasma edge should therefore remain relatively unimportant. Particle tracking simulations are necessary to further elucidate the behavior of electrons near the very plasma edge.
\section{Conclusions \label{sec7}}
In this paper, we have studied the ponderomotive forces induced in a subwavelength plasma by an externally applied electromagnetic wave. We found that the ponderomotive force in the plasma bulk is directed outwards for natural profiles $dn_e/dr<0$ and inwards for 'inverted' profiles $dn_e/dr>0$. For a completely homogeneous plasma, a spherically symmetric compressive ponderomotive force remains, suggesting possibilities for contactless ponderomotive manipulation of homogeneous subwavelength objects. Furthermore, we showed that the force in the plasma bulk is accompanied by a strongly peaked outward ponderomotive force near sharp plasma edges. In the limit that the plasma boundary tends to a discontinuous step in the density, this force peak tends to a ponderomotive surface force, which in turn makes an essential contribution to the total radiation pressure on the plasma. Finally, we have discussed the feasibility of radiative acceleration of ultracold plasmas. Based on existing technologies and conservative estimates, we estimated that these plasmas may be accelerated to keV ion energies, resulting in a neutralized beam with a brightness comparable to current high-performance ion sources. Subsequent fluid simulations should address the plasma dynamics and the self-consistent evolution of the density profile, while particle tracking simulations may identify departures from the continuum description adopted in this paper, especially concerning particles near the plasma edge. Extension of our results to plasma sizes comparable to or larger than the wavelength will be very interesting as well. It is clear that ponderomotive forces play an important role in electromagnetically driven finite-sized plasmas in general, and in ultracold plasmas in particular. A thorough understanding of these forces will enable opportunities for active ponderomotive plasma manipulation, including the compression and acceleration of ultracold neutral plasmas.
\begin{acknowledgments}
This work is part of the research program of FOM, which is financially supported by NWO.
\end{acknowledgments}
\appendix
\section{Quasistatic limit from general field expressions \label{secA}}
We first estimate which potential Eq. (\ref{4}) becomes dominant in the quasistatic limit. As mentioned in \ref{sec2c}, $f_n^{e,m}\sim(kb)^n$ if $kb\ll1$. Consequently, the lowest-order modes $\Pi^{e,m}_1$ are dominant, the high-order modes being progressively smaller. Furthermore, assuming in Eq. (\ref{2}) that symbolically $\nabla\sim b^{-1}$, it follows that $|\bm{E}^e_n|\gg|\bm{E}^m_n|$. Hence, the dominant contribution to the electric field is the electric dipole mode, which is equal to
\begin{align}
\bm{E}\approx\bm{E}^e_1\approx E_0\nabla\left(\frac{3}{2k\epsilon}\frac{d(rf_1^e)}{dr}\sin\theta\cos\varphi \right)\label{A1}.
\end{align}
Here, the identity
\begin{align}
\frac{1}{k\epsilon}\nabla\times\left(\bm{r}\times\nabla\Pi^e\right)=-\frac{1}{k}\nabla\left(\frac{1}{\epsilon}\frac{\partial(r\Pi^e)}{\partial r}\right)-k\bm{r}\Pi^e\label{A1a}
\end{align}
has been used. Comparison of Eqs. (\ref{17}) and (\ref{A1}) shows that the function $\xi=-(3/2k\epsilon)d(rf_1^e)/dr$ must reduce to $\psi$ in the quasistatic limit, the latter being defined by the boundary value problem (\ref{18})-(\ref{21}). This can be shown by noting that in Eq. (\ref{5}) the propagation term $k^2\epsilon$ is much smaller than the other terms in the quasistatic limit. Neglecting the propagation term, taking $n=1$, and multiplying Eq. (\ref{5}) by $d/dr+2/r$, yields
\begin{align}
0=\left[\frac{d^2}{dr^2}+\left(\frac{2}{r}+\frac{1}{\epsilon}\frac{d\epsilon}{dr}\right)\frac{d}{dr}-\frac{2}{r^2}\right]\xi.\label{A2}
\end{align}
Similarly, multiplying Eq. (\ref{22}) by $-3/kb$ and taking $n=1$, approximating the Bessel functions by their limiting value for small argument, and rewriting $f_1^e(b)$ using Eq. (\ref{5}), gives
\begin{align}
-3=\left(\epsilon\frac{d\xi}{dr}+\frac{2\xi}{r}\right)_{r=b}.\label{A3}
\end{align}
From Eqs. (\ref{18}), (\ref{21}), (\ref{A2}) and (\ref{A3}), $\psi$ and $\xi$ satisfy the same differential equation and the same boundary conditions, which shows that $\xi\approx\psi$ when $kb\ll1$. Hence the general solution for the electric field given in the section \ref{sec2a} approaches the quasistatic field given in section \ref{sec2b}.

\clearpage
\begin{widetext}
\section{Explicit expressions for ponderomotive forces \label{secB}}
In the ponderomotive volume force density Eq. (\ref{32}),
\begin{align}
R^{r1}_{nm}&=\frac{d(rR^{\theta1}_{nm})}{dr}=\frac{2-\delta_{nm}}{k}\left(g^m_n\frac{dg^m_m}{dr}+g^m_m\frac{dg^m_n}{dr}\right);\label{B1}\\
R^{r2}_{nm}&=\frac{d(rR^{\theta2}_{nm})}{dr}=n(n+1)m(m+1)\frac{2-\delta_{nm}}{(kr)^3\epsilon^2}\left[g_n^e\frac{d(rg_m^e)}{dr}+g_m^e\frac{d(rg_n^e)}{dr}-2\left(2+\frac{r}{\epsilon}\frac{d\epsilon}{dr}\right)g^e_ng^e_m\right];\label{B2}\\
R^{r3}_{nm}&=\frac{d(rR^{\theta3}_{nm})}{dr}=\frac{2-\delta_{nm}}{(kr)^3\epsilon^2}\!\left\{\!\left[n(n\!+\!1)-\epsilon(kr)^2\right]g^e_n\frac{d(rg^e_m)}{dr}+\!\left[m(m\!+\!1)-\epsilon(kr)^2\right]g^e_m\frac{d(rg^e_n)}{dr}-2\frac{d(rg^e_n)}{dr}\frac{d(rg^e_m)}{dr}\right\};\label{B3}\\
R^{r4}_{nm}&=\frac{d(rR^{\theta4}_{nm})}{dr}=\frac{2}{(kr)^2\epsilon}\left\{\frac{d(rg^e_n)}{dr}\frac{d(rg_m^m)}{dr}-2g^m_m\frac{d(rg^e_n)}{dr}+\left[n(n+1)-\epsilon(kr)^2\right]g^e_ng^m_m\right\};\label{B4}\\
R^{\theta1}_{nm}&=\makebox[15mm][c]{$R^{\varphi1}_{n,m}$}=\frac{2-\delta_{nm}}{kr}\,g^m_ng^m_m;\label{B5}\\
R^{\theta2}_{nm}&=\makebox[15mm][c]{$R^{\varphi1}_{n,m}$}=n(n+1)m(m+1)\frac{2-\delta_{nm}}{(kr)^3\epsilon^2}g^e_ng^e_m;\label{B6}\\
R^{\theta3}_{nm}&=\makebox[15mm][c]{$R^{\varphi1}_{n,m}$}=\frac{2-\delta_{nm}}{(kr)^3\epsilon^2}\frac{d(rg^e_n)}{dr}\frac{d(rg^e_m)}{dr};\label{B7}\\
R^{\theta4}_{nm}&=\makebox[15mm][c]{$R^{\varphi1}_{n,m}$}=\frac{2}{(kr)^2\epsilon}\,g^m_m\frac{d(rg^e_n)}{dr};\label{B8}\\
S^{r1}_{nm}&=\frac{(2n+1)(2m+1)}{n(n+1)m(m+1)}\left(\frac{P_n^1P_m^1}{\sin^2\theta}\cos^2\varphi+\frac{dP_n^1}{d\theta}\frac{dP_m^1}{d\theta}\sin^2\varphi\right);\label{B9}\\
S^{r2}_{nm}&=\frac{(2n+1)(2m+1)}{n(n+1)m(m+1)}P_n^1P_m^1\cos^2\varphi;\label{B10}\\
S^{r3}_{nm}&=\frac{(2n+1)(2m+1)}{n(n+1)m(m+1)}\left(\frac{dP_n^1}{d\theta}\frac{dP_m^1}{d\theta}\cos^2\varphi+\frac{P_n^1P_m^1}{\sin^2\theta}\sin^2\varphi\right);\label{B11}\\
S^{r4}_{nm}&=\frac{(2n+1)(2m+1)}{n(n+1)m(m+1)}\left(\frac{dP_n^1}{d\theta}\frac{P_m^1}{\sin\theta}\cos^2\varphi+\frac{P^1_n}{\sin\theta}\frac{dP_m^1}{d\theta}\sin^2\varphi\right);\label{B12}\\
S^{\theta j}_{nm}&=\frac{\partial S^{rj}_{nm}}{\partial\theta};\label{B13}\\
S^{\varphi j}_{nm}&=\frac{1}{\sin\theta}\frac{\partial S^{rj}_{nm}}{\partial\varphi},\label{B14}
\end{align}
where $\delta_{nm}$ is the Kronecker delta, and $j=1\dots4$. In Eqs. (\ref{B3})-(\ref{B4}), the differential equation (\ref{22a}) has been applied to rewrite second derivatives.
\end{widetext}
The $z$ component $f_z$ of Eq. (\ref{32}) consists of terms that are proportional to $X^j_{nm}=R^{rj}_{nm}S^{rj}_{nm}\cos\theta-R^{\theta j}_{nm}S^{\theta j}_{nm}\sin\theta$, with $j=1\dots4$. In the volume integration of $f_z$ in Eq. (\ref{38}), integrating by parts the second term of $X^j_{nm}$ with respect to $\theta$, and using the functional relations in Eqs. (\ref{B1})-(\ref{B4}) and (\ref{B13}), transforms the angular integrations to
\begin{align}
\int X^j_{nm}\,d\Omega&=\frac{d(r^3R^{\theta j}_{nm})}{dr}\int S^{rj}_{nm}\cos\theta d\Omega.\label{B15}
\end{align}
The remaining four integrals $j=1\dots4$ on the right side of Eq. (\ref{B15}) are equal to \footnote{Eq. (\ref{B17}) is obtained by writing $(2m+1)P_m^1=mP_{m+1}^1+(m+1)P_{m-1}^1$ and using the orthogonality relations of the Legendre functions \cite{Abramowitz}; Eqs. (\ref{B16}) and (\ref{B19}) are well-known from scattering theory in calculation of the asymmetry parameter \cite{Bohren}.}
\begin{align}
\hspace{-2mm}\int S^{r1}_{nm}\cos\theta d\Omega&=\left\{\!\!\makebox[2.5cm][l]{\setlength{\extrarowheight}{5pt}$\begin{array}{l}\frac{2\pi q^2(q+1)(q+2)^2}{(2q+1)(2q+3)}\\0\end{array}$}\setlength{\extrarowheight}{5pt}\begin{array}{l}\vphantom{\frac{2q^2(q+1)(q+2)^2}{(2q+1)(2q+3)}}m=n\pm1;\\m\neq n\pm1;\end{array}\right.\label{B16}\\
\hspace{-2mm}\int S^{r2}_{nm}\cos\theta d\Omega&=\left\{\!\!\makebox[2.5cm][l]{\setlength{\extrarowheight}{5pt}$\begin{array}{l}\frac{2\pi q(q+1)(q+2)}{(2q+1)(2q+3)}\\0\end{array}$}\setlength{\extrarowheight}{5pt}\begin{array}{l}\vphantom{\frac{2q(q+1)(q+2)}{(2q+1)(2q+3)}}m=n\pm1;\\m\neq n\pm1;\end{array}\right.\label{B17}\\
\hspace{-2mm}\int S^{r3}_{nm}\cos\theta d\Omega&=\int S^{r1}_{n,m}\cos\theta d\Omega;\label{B18}\\
\hspace{-2mm}\int S^{r4}_{nm}\cos\theta d\Omega&=\left\{\!\!\makebox[2.5cm][l]{\setlength{\extrarowheight}{5pt}$\begin{array}{l}\frac{2\pi n(n+1)}{2n+1}\\0\end{array}$}\setlength{\extrarowheight}{5pt}\begin{array}{l}\vphantom{\frac{2n(n+1)}{2n+1}}m=n;\\m\neq n,\end{array}\right.\label{B19}
\end{align}
with $q=\min(n,m)$. The resulting total volume force is given in Eq. (\ref{33}), in which
\begin{align}
\hspace{-2.3mm}Y^1_n&=\frac{n(n+2)}{n+1}\!\!\int_0^{b^-}\hspace{-3mm}\chi\frac{d}{dr}\!\left[\vphantom{\frac{1}{\epsilon^2}}(kr)^2g_n^mg_{n+1}^m\right]\!dr;\label{B20}\\
\hspace{-2.3mm}Y^2_n&=n(n+1)(n+2)\!\!\int_0^{b^-}\hspace{-3mm}\chi\frac{d}{dr}\!\left[\frac{g_n^eg_{n+1}^e}{\epsilon^2}\right]\!dr;\label{B21}\\
\hspace{-2.3mm}Y^3_n&=\frac{n(n+2)}{n+1}\!\!\int_0^{b^-}\hspace{-3.5mm}\chi\frac{d}{dr}\!\left[\frac{1}{\epsilon^2}\frac{d(rg_n^e)}{dr}\frac{d(rg_{n+1}^e)}{dr}\right]\!dr;\label{B22}\\
\hspace{-2.3mm}Y^4_n&=\frac{2n+1}{n(n+1)}\!\int_0^{b^-}\hspace{-3mm}\chi\frac{d}{dr}\!\left[\frac{krg_n^m}{\epsilon}\frac{d(rg_n^e)}{dr}\right]\!dr.\label{B23}
\end{align}
\section{Radial functions for homogeneous sphere \label{secC}}
Solving Eqs. (\ref{22a})-(\ref{28}) for a homogeneous sphere with permittivity $\epsilon_1$ gives $g_n^{e,m}=A_n^{e,m}j_n(\sqrt{\epsilon_1}kr)$, with
\begin{align}
A^{e,m}_n&=\frac{y_n}{kb\raisebox{2pt}{\big|}h_n^{(1)}\raisebox{2pt}{\big|}^2G^{e,m}_n}\label{C1}\\
G^{e,m}_n&\equiv\delta^{e,m}\!\left(n\widetilde{\jmath}_{n}\!-\!\sqrt{\epsilon_1}kb\widetilde{\jmath}_{n-1}\right)\!+\!\left(1+b\frac{d}{dr}\ln\left|h_n^{(1)}\right|\right)\widetilde{\jmath}_n,\nonumber
\end{align}
where $\widetilde{\jmath}_n$ denotes the spherical Bessel function with argument $\sqrt{\epsilon_1}\,kb$ and $j_n,y_n,h^{(1)}_n$ are spherical Bessel functions with argument $kb$. Substituting $g^{e,m}_n$ in Eq. (\ref{30}) and expanding braces yields
\begin{align}
\gamma^{e,m}_n=\frac{kb\raisebox{2pt}{\big|}h_n^{(1)}\raisebox{2pt}{\big|}^2h_n^{(2)}G^{e,m}_n}{y_n\left[\widetilde{\jmath}_n-ikb\raisebox{2pt}{\big|}h_n^{(1)}\raisebox{2pt}{\big|}^2G^{e,m}_n\right]},\label{C2}
\end{align}
with $h^{(2)}_n$ the $n$th-order spherical Hankel function of the second kind \cite{Abramowitz} and argument $kb$. Multiplying in Eq. (\ref{C2}) the term $\widetilde{\jmath}_n$ by the identity $1=(j_{n+1}y_n-j_ny_{n+1})(kb)^2$, and simplifying the denominator, gives
\begin{align}
A^e_n\gamma^e_n&=\frac{i(kb)^{-2}}{\sqrt{\epsilon_1}h_n^{(1)}\widetilde{\jmath}_{n+1}-h_{n+1}^{(1)}\widetilde{\jmath}_n};\label{C3}\\[2pt]
A^m_n\gamma^m_n&=\frac{i(kb)^{-2}}{\dfrac{h_n^{(1)}\widetilde{\jmath}_{n+1}}{\sqrt{\epsilon_1}}-h_{n+1}^{(1)}\widetilde{\jmath}_n+\dfrac{(n+1)\chi_1}{\epsilon_1kb}h_n^{(1)}\widetilde{\jmath}_n}.\label{C4}
\end{align}
Eqs. (\ref{C3})-(\ref{C4}) are equal to $c_n$ and $\sqrt{\epsilon_1}d_n$ respectively, where $c_n$ and $d_n$ are the internal Mie coefficients \cite{Bohren}. Taylor expansions about $kb=0$ of Eqs. (\ref{C1})-(\ref{C2}) are
\begin{align}
A^e_1&=\frac{3\sqrt{\epsilon_1}}{\epsilon_1+2}\left(1+\frac{\chi_1(\epsilon_1+10)}{10(\epsilon_1+2)}(kb)^2+\dots\right);\label{C5}\\
A^e_2&=\frac{5}{3(2\epsilon_1+3)}\left(1+\frac{\chi_1(2\epsilon_1+7)}{14(2\epsilon_1+3)}(kb)^2+\dots\right);\label{C6}\\
A^e_3&=\frac{7\epsilon_1^{-1/2}}{(3\epsilon_1+4)}\left(1+\frac{\chi_1(5\epsilon_1+12)}{30(3\epsilon_1+4)}(kb)^2+\dots\right);\label{C6a}\\
A^m_1&=\frac{1}{\sqrt{\epsilon_1}}\left(1+\frac{\chi_1}{6}(kb)^2+\dots\right);\label{C7}\\
A^m_2&=\frac{1}{\epsilon_1}\left(1+\frac{\chi_1}{10}(kb)^2+\dots\right);\label{C7a}\\
\gamma^e_1&=1+\frac{2i\chi_1(kb)^3}{3(\epsilon_1+2)}\!\left(1\!+\frac{3(\epsilon_1-2)}{5(\epsilon_1+2)}(kb)^2\!+\dots\right);\label{C8}\\
\gamma^e_2&=1+\frac{i\chi_1(kb)^5}{15(2\epsilon_1+3)}+\dots;\label{C9}\\
\gamma^m_1&=1+\frac{i\chi_1(kb)^5}{45}+\dots\,\,.\label{C10}
\end{align}
The imaginary part of other $\gamma^{e,m}_n$ are of order $O[(kb)^7]$.
\end{document}